\documentclass[twocolumn,secnumarabic,amssymb, nobibnotes, aps, prl, floatfix, superscriptaddress]{revtex4-1}

\usepackage[utf8]{inp utenc}

 \usepackage{graphicx}
\usepackage{bm}
\usepackage{amsmath} \usepackage{braket} \usepackage{epsfig} 
\usepackage{tensor}
\usepackage{CJKutf8}
\usepackage[version=4]{mhchem}
\usepackage{titlesec}
\usepackage{ifthen}
\usepackage[misc]{ifsym}
\usepackage{sidecap}
\usepackage{listings}
\usepackage[para,online,flushleft]{threeparttablex}

\usepackage{hyperref}
\usepackage[all]{hypcap}
\usepackage{float}

\usepackage{caption,subcaption}
\DeclareCaptionLabelSeparator{vert}{ $\boldsymbol{\vert}$ }
\usepackage{multibib}
\newcites{meth}{References}

\usepackage{xcolor}
\definecolor{pastelgray}{rgb}{0.81, 0.81, 0.77}
\definecolor{beaublue}{rgb}{0.9, 0.9, 0.93}

\newcommand{\Pbbold}{\ensuremath{\boldsymbol{{}^{208}\mathrm{Pb}}}}
\newcommand{\Pb}{\ensuremath{{}^{208}{\rm Pb}}}
\newcommand{\Ca}{\ensuremath{{}^{48}{\rm Ca}}}

\newcommand{\Ox}{\ensuremath{{}^{16}{\rm O}}}
\newcommand{\rskinPbbold}{\ensuremath{\boldmath{R_\text{skin}(\Pb)}}}
\newcommand{\rskinPb}{\ensuremath{R_\text{skin}(\Pb)}}
\newcommand{\rskinCa}{\ensuremath{R_\text{skin}(\Ca)}}

\renewcommand{\figurename}{Figure}

\setcitestyle{super}

\titleformat{\section}{\normalsize\raggedright\bfseries\fontsize{10}{12}}{\arabic{section}.}{1em}{}
\titleformat{\subsection}{\small\raggedright\bfseries}{\arabic{section}.}{1em}{}

\titlespacing\section{0pt}{12pt plus 4pt minus 2pt}{0pt plus 2pt minus 2pt}

\begin{document}
\begin{CJK*}{UTF8}{gbsn}

\begin{titlepage}
  {\fontsize{26}{10}
    \textbf{\textcolor{black}{\flushleft 
    \emph{Ab initio} predictions link the neutron skin of $^{208}$Pb to nuclear forces}}}\\
{
  Baishan Hu$^{1,*}$,
  Weiguang Jiang$^{2,*}$,
  Takayuki Miyagi$^{1,3,4,*}$,
  Zhonghao Sun$^{5,6,*}$,
  Andreas Ekstr{\"o}m$^{2}$,
  Christian Forss{\'e}n$^{2,\textrm{\Letter}}$,
  Gaute Hagen$^{6,5,1}$,
  Jason D. Holt$^{1,7}$,
  Thomas Papenbrock$^{5,6}$,
  S. Ragnar Stroberg$^{8,9}$,
  Ian Vernon$^{10}$
}

{
\fontsize{6}{10}{
\selectfont
$^{1}$TRIUMF, 4004 Wesbrook Mall, Vancouver, BC V6T 2A3, Canada.
$^{2}$Department of Physics, Chalmers University of Technology, SE-412 96 G{\"o}teborg, Sweden.
$^{3}$Technische Universit\"at Darmstadt, Department of Physics, 64289 Darmstadt, Germany.
$^{4}$ExtreMe Matter Institute EMMI, GSI Helmholtzzentrum f\"ur Schwerionenforschung GmbH, 64291 Darmstadt, Germany.
$^{5}$Department of Physics and Astronomy, University of Tennessee, Knoxville, Tennessee 37996, USA.
$^{6}$Physics Division, Oak Ridge National Laboratory, Oak Ridge, Tennessee 37831, USA.
$^{7}$Department of Physics, McGill University, 3600 Rue University, Montr{\'e}al, QC H3A 2T8, Canada.
$^{8}$Department of Physics, University of Washington, Seattle, Washington 98195, USA.
$^9$Physics Division, Argonne National Laboratory, Lemont, Illinois 60439, USA.
$^{10}$Department of Mathematical Sciences, Durham University, Stockton Road, Durham, DH1 3LE, UK.
$^*$contributed equally.
$^\textrm{\Letter}$corresponding author:  christian.forssen@chalmers.se
}}

\end{titlepage}
\end{CJK*}

{\bf 
Heavy  atomic nuclei have an excess of neutrons over protons, which leads to the formation of a neutron skin whose thickness is sensitive to details of the nuclear force.   
%
This links atomic nuclei to properties of neutron stars, thereby relating objects that differ in size by orders of magnitude. The nucleus \Pbbold{} is of particular interest because it exhibits a simple structure and is experimentally accessible. However, computing such a heavy nucleus has been out of reach for \textit{\bfseries ab initio} theory.
%
By combining advances in quantum many-body methods, statistical tools, and emulator technology, we make quantitative predictions for the properties of \Pbbold{} starting from nuclear forces that are consistent with symmetries of low-energy quantum chromodynamics. 
We explore $\boldsymbol{10^9}$ different nuclear-force parameterisations via history matching, confront them with data in select light nuclei, and arrive at an importance-weighted ensemble of interactions. 
%
We accurately reproduce bulk properties of \Pbbold{} and determine the neutron skin thickness, which is smaller and more precise than a recent 
extraction from parity-violating electron scattering but in agreement with other experimental probes. 
%
%
This work demonstrates how realistic two- and three-nucleon forces act in a heavy nucleus and allows us to make quantitative predictions across the nuclear landscape.
}

%
Neutron stars are extreme astrophysical objects whose interiors may contain exotic new forms of matter.
The structure and size of neutron stars are linked to the thickness of neutron skins in atomic nuclei via the neutron-matter equation of state~\cite{brown2000,horowitz2001, essick2021}.
The nucleus \Pb{} is an attractive target for exploring this link in both experimental~\cite{tarbert2013,adhikari2021} and theoretical~\cite{tsang2012, horowitz2001, roca-maza2011} studies, due to the large excess of neutrons and its simple structure.
Mean-field calculations predict a wide range for $\rskinPb{}$ because the isovector parts of nuclear energy density functionals are not well constrained by binding energies and charge radii~\cite{pethick1995,brown1998,horowitz2001,roca-maza2011}.
Additional constraints may be obtained~\cite{reinhard2021} by including the electric dipole polarisability of \Pb{}, though this comes with a model dependence~\cite{piekarewicz2012} which is difficult to quantify.
In general, estimation of systematic theoretical uncertainties is  a challenge for mean-field theory.
%

\begin{figure}[bht]
  \includegraphics[width=1.0\linewidth]{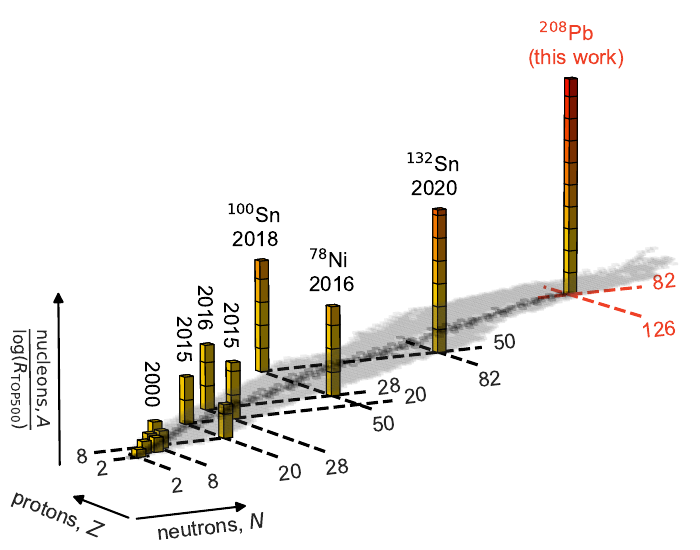}
  \caption{ \textbf{Trend of realistic \emph{ab initio}  computations for the nuclear $A$-body problem.}
    The bars highlight years of first realistic computations of doubly magic nuclei. The height of each bar
    corresponds to the mass number $A$ divided by the logarithm of the
    total compute power $R_\mathrm{TOP500}$
    (in flop/second) of the pertinent TOP500 list~\cite{top500}.
    This ratio would be approximately constant if progress were solely due to exponentially increasing computing power.
    However, 
    algorithms
    which instead scale polynomially in $A$ have greatly increased the
    reach. 
    \label{fig:abinitio}
  }
\end{figure}

In contrast, precise {\it ab initio} computations, which provide a path to comprehensive uncertainty estimation, have been accomplished for the neutron-matter equation of state~\cite{hagen2014,tews2016,drischler2020} and the neutron skin in the medium-mass nucleus $^{48}$Ca~\cite{hagen2015}.
But up to now treating \Pb{} within the same framework was out of reach.
%
%
Due to breakthrough developments in quantum many-body methods, such computations are now becoming feasible for heavy nuclei~\cite{morris2018,arthuis2020,stroberg2021,miyagi2021}.
The {\it ab initio} computation of \Pb{}
we report here
represents a
significant step in mass number from the previously computed tin
isotopes~\cite{morris2018,arthuis2020}, as illustrated in
Figure~\ref{fig:abinitio}.
 The complementary statistical analysis in this work is enabled by
 emulators (for mass number $A\le 16$) which mimic the outputs of many-body solvers, but are orders of magnitude faster.

In this paper we develop a unified \emph{ab initio} framework to
link the physics of nucleon-nucleon scattering and few-nucleon systems to
properties of medium- and heavy-mass nuclei up to 
\Pb{}, and ultimately to the nuclear matter equation of state near saturation density. 

\section{Linking models to reality}

Our approach to constructing nuclear interactions is based on chiral effective
field theory (EFT)~\cite{epelbaum2009,machleidt2011,Hammer:2019poc}. In this theory
the long-range part of the strong nuclear force is known and stems
from pion exchanges, while the unknown short-range contributions are
represented as contact interactions; we also include the $\Delta$ isobar degree of
freedom~\cite{ordonez1996}. At next-to-next-to leading order in Weinberg's power
counting, the four pion-nucleon low-energy constants (LECs) are 
tightly fixed from pion-nucleon scattering data\cite{hoferichter2015}.
The 13 additional LECs in the nuclear potential must be constrained from data.

We use history matching~\cite{Vernon:2010,Vernon:2014} to explore the
modeling capabilities of \emph{ab initio} methods by
identifying a \emph{non-implausible} region in the vast parameter space
of LECs, for which the model output 
yields
acceptable agreement with selected low-energy experimental data---here denoted
history-matching observables.
%
%
The key to efficiently analyze this high-dimensional 
parameter space is the use of emulators based on eigenvector
continuation~\cite{frame2018,Konig:2019adq,ekstrom2019} that accurately mimic the outputs of 
the \emph{ab initio} methods at several orders of magnitude lower
computational cost.
We consider the following history-matching observables:
nucleon-nucleon scattering phase shifts up to an energy of 200 MeV; the energy, radius, and quadrupole moment of ${}^{2}\text{H}$; and
the energies and radii of ${}^{3}\text{H}$, ${}^{4}\text{He}$, and
\Ox{}. 
We perform five waves of
this global parameter search---see Extended Data Figures~\ref{fig:hm_waves} and \ref{fig:NI_phase_shifts}---sequentially ruling out implausible
LECs that yield model predictions too far from
experimental data.
For this purpose we use an implausibility measure (see Methods) that links our model predictions and experimental observations as
\begin{equation}
  z =  M (\theta) + \varepsilon_\mathrm{exp}+ \varepsilon_\mathrm{em}+ \varepsilon_\mathrm{method} +
  \varepsilon_\mathrm{model}.
  \label{eq:modelreality}
\end{equation}
Here, experimental observations, $z$, are related to emulated \emph{ab initio} predictions $M(\theta)$
via random variables $\varepsilon_\mathrm{exp}$, $\varepsilon_\mathrm{em}$, $\varepsilon_\mathrm{method}$,
$\varepsilon_\mathrm{model}$ that represent experimental
uncertainties, emulator precision, 
method approximation errors, and the model discrepancy due to the EFT truncation at next-to-next-to leading order,
respectively. The parameter vector $\theta$ corresponds to the 17 LECs at this order.
The method error represents, e.g., model-space
truncations and other approximations in the employed \emph{ab initio} many-body
solvers. The model
discrepancy $\varepsilon_\mathrm{model}$ can be probabilistically specified since we assume to operate with an order-by-order improvable EFT description of the nuclear interaction (see Methods for
details).

The final result of the five history-matching waves is a set of 34 non-implausible samples in the 17-dimensional parameter space of the LECs. We then perform \emph{ab initio} calculations for nuclear observables in \Ca{} and \Pb{}, as well as for properties of infinite nuclear matter. 

\section{\emph{Ab initio} computations of \Pb{}}
%
We employ the coupled-cluster
(CC)~\cite{kuemmel1978,bartlett2007,hagen2014}, the in-medium
similarity renormalization group
(IMSRG)~\cite{hergert2016} and many-body perturbation theory (MBPT) methods to approximately solve the Schr\"odinger equation and obtain the ground-state
energy and nucleon densities of \Ca{} and \Pb{}. We analyze the
model-space convergence and use the differences between CC, IMSRG and MBPT results
to estimate the method approximation errors, see Methods and Extended Data
Figures~\ref{fig:PbConvergence} and \ref{fig:spcc_method_prec}.
The computational cost of these methods scales (only) polynomially with increasing numbers of nucleons and single-particle orbitals.
\begin{figure}[!htb]
  \includegraphics[width=0.8\linewidth]{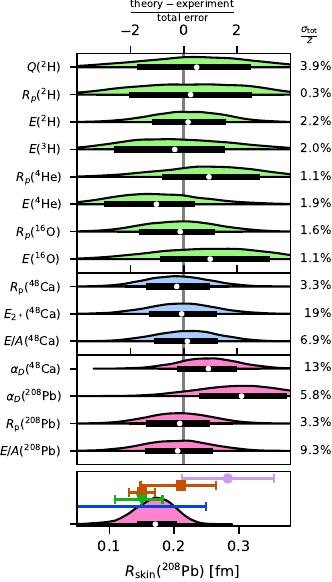}
  \caption{ \textbf{\emph{Ab initio}  posterior predictive distributions for light to
      heavy nuclei.}
    Model checking is indicated by green (blue) distributions,
    corresponding to observables used for history-matching (likelihood calibration),
   while pure predictions are shown as pink distributions. 
    Nuclear observables shown are: quadrupole moment $Q$, point-proton radii $R_\mathrm{p}$, ground-state energies $E$ (or energy per nucleon $E/A$), $2^+$ excitation energy $E_{2^+}$, and electric dipole polarizabilities $\alpha_D$.
      See 
      Extended Data Table~\ref{tab:error_assignments} for numerical specification
      of experimental data ($z$), errors
      ($\sigma_i$), medians (white circle) and
      68\% credibility regions (thick bar). The
      prediction for \rskinPb{} in the bottom 
      panel is shown in an absolute scale and compared to experimental
      results using electroweak~\cite{adhikari2021} (purple),
      hadronic~\cite{Trzcinska:2001sy,zenihiro2010} (red),
      electromagnetic~\cite{tarbert2013} (green), and gravitational
      waves~\cite{Fattoyev:2017jql} (blue) probes (from top to bottom; see Extended Data Figure~\ref{fig:1S0_and_L_vs_Rskin}b for details).
    \label{fig:blobbogram}
  }
\end{figure}
The main challenge in computing \Pb{} is the vast number of matrix
elements of the three-nucleon force which must be handled. 
We overcome this limitation by using a recently introduced storage
scheme in which we only store linear combinations of matrix
elements directly entering the normal-ordered two-body
approximation~\cite{miyagi2021} (see Methods for details).

Our \emph{ab initio} predictions for finite nuclei are summarized in
Figure~\ref{fig:blobbogram}. The statistical approach that leads to these results is
composed of three stages. 
First, history matching identified a set of 34 
non-implausible interaction parametrizations.
Second, model calibration is performed by weighting these parametrizations---serving
as prior samples---using a likelihood measure according to the principles of sampling/importance resampling\cite{Smith:1992aa}. This yields 34 weighted
samples from the LEC posterior probability density function, see Extended Data Figure~\ref{fig:importanceweights}. Specifically we assume independent EFT
and many-body method errors and construct a normally distributed data-likelihood encompassing the ground-state energy per nucleon $E/A$ and
point-proton radius $R_\mathrm{p}$ for \Ca{}, and the
energy $E_{2^+}$ of its first excited $2^+$ state.
Our final predictions are therefore conditional on this calibration data.

We have tested the
sensitivity of final results to the likelihood definition by repeating
the calibration with a
non-diagonal covariance matrix or a Student-t distribution with
heavier tails, finding small ($\sim 1\%$) differences in the predicted
credible regions.
The EFT truncation errors are quantified by studying \emph{ab initio}
predictions at different orders in the power counting for
\Ca{} and infinite nuclear matter.
We validate our \emph{ab initio} model and error assignments by computing the posterior predictive distributions, including all relevant sources of uncertainty, for both the replicated calibration data (blue colour) and the history-matching observables (green colour), see Figure~\ref{fig:blobbogram}. The percentage ratios
$\sigma_\text{tot}/z$ of the (theory dominated)
total uncertainty to the experimental value are given
in the right margin.

Finally, having built confidence in our \emph{ab initio} model and
underlying assumptions, we predict \rskinPb{}, $E/A$ and
$R_\mathrm{p}$ for \Pb{}, $\alpha_D$ for \Ca{} and \Pb{} as well as
nuclear matter properties, by employing  importance resampling~\cite{Smith:1992aa}.
The corresponding posterior predictive distributions are shown in the lower panels of
Figure~\ref{fig:blobbogram} (pink colour).
Our prediction $\rskinPb{} = 0.14 - 0.20$~fm exhibits a mild
  tension with the value extracted from the recent parity-violating
electron scattering experiment PREX~\cite{adhikari2021} but is
consistent with the skin thickness extracted from
elastic proton scattering~\cite{zenihiro2010}, antiprotonic atoms~\cite{Trzcinska:2001sy} and coherent pion photoproduction~\cite{tarbert2013} as well as constraints from
gravitational waves from merging neutron
stars~\cite{Fattoyev:2017jql}.

We also compute the weak form factor 
$F_\mathrm{w}(Q^2)$ at momentum transfer $Q_\mathrm{PREX}=0.3978(16)$~fm$^{-1}$,
which is more directly related to the parity-violating asymmetry measured in the PREX experiment.
We observe a strong correlation with the more precisely measured electric charge form
factor $F_\mathrm{ch}(Q^2)$, as shown in Figure~\ref{fig:ppd_FW}b. While we have not
quantified the EFT and method errors for these observables, we find
a small variance among the 34 non-implausible predictions for the
difference $F_\mathrm{w}(Q^2) - F_\mathrm{ch}(Q^2)$ for both
\Ca{} and \Pb{} as shown in Figure~\ref{fig:ppd_FW}c.

\section{\emph{Ab initio} computations of infinite nuclear matter}
%
\begin{figure*}[!htb]
\includegraphics[width=0.94\linewidth]{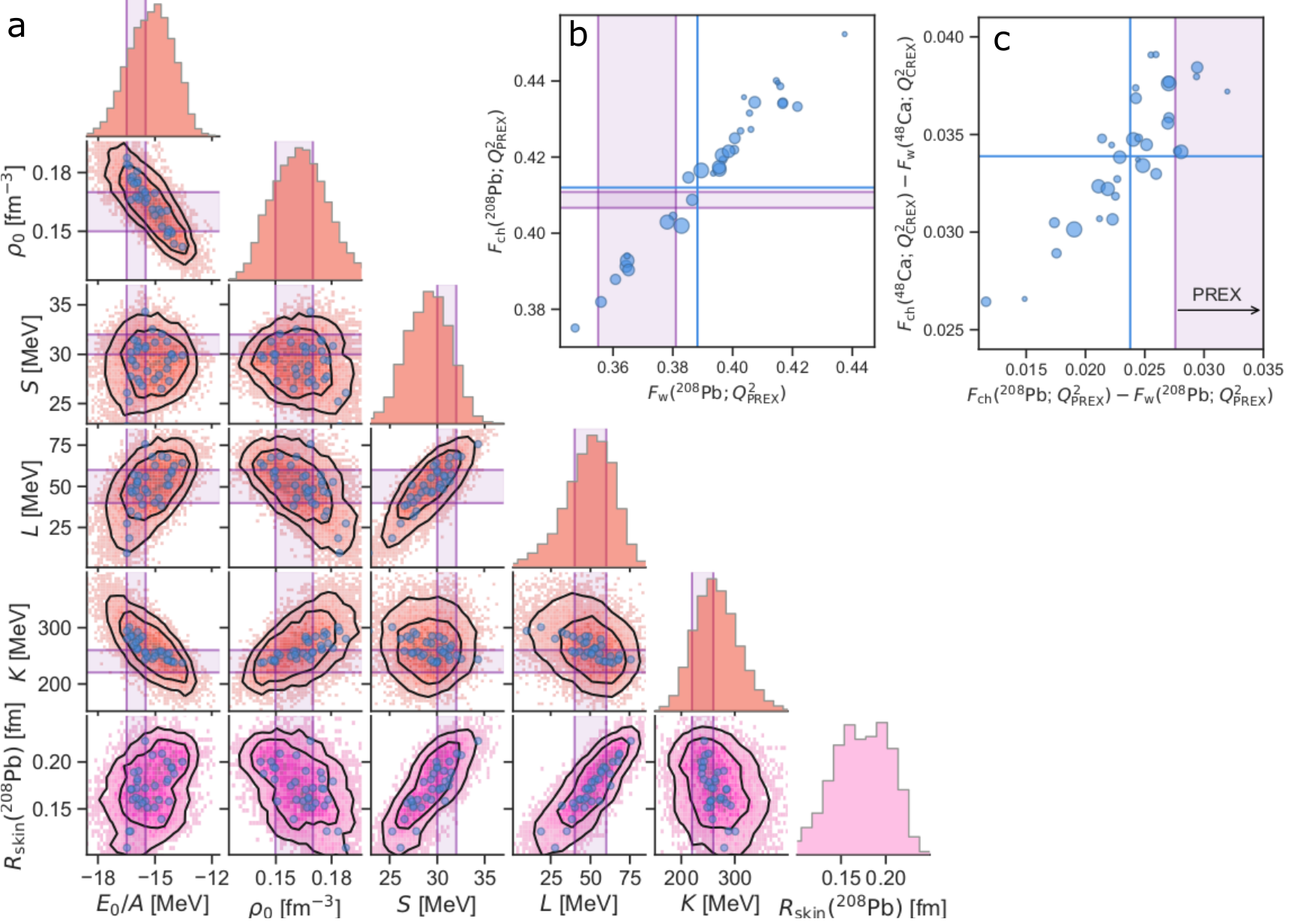}
\caption[Posterior predictive distribution]
{\textbf{Posterior predictive distribution for \rskinPb{}
    and nuclear matter at saturation density.}
    \textbf{a}, Predictions for the saturation energy per particle
    $E_0/A$ and density $\rho_0$ of
    symmetric nuclear matter, its compressibility $K$, the symmetry
    energy $S$, and its slope $L$ are
    correlated with the those for \rskinPb{}.
    The bivariate distribution include 68\% and 90\%
    credible regions (black lines) and a scatter plot of the
    predictions with the 34 non-implausible samples before error
    sampling. Empirical nuclear-matter properties are indicated by
    purple bands (see Extended Data Table~\ref{tab:ppd}). 
    \textbf{b}, Predictions with the 34 non-implausible samples for
    the electric $F_\mathrm{ch}$ versus weak $F_\mathrm{w}$ charge form factors for 
    \Pb{} at the momentum transfer 
    considered in the PREX experiment~\cite{adhikari2021}. 
  \textbf{c}, Difference between electric and weak charge form factors
  for \Ca{} and \Pb{} at the momentum transfers $Q_\mathrm{CREX} =
  0.873$~fm$^{-1}$  and $Q_\mathrm{PREX} = 0.3978$~fm$^{-1}$ that are relevant for
  the CREX and PREX experiments, respectively. 
  Experimental data (purple bands) in panels b and c are from Ref.~\cite{adhikari2021}, the size of the markers indicate the importance weight, and blue lines correspond to weighted means. 
  \label{fig:ppd_FW}
}
\end{figure*}
We also make predictions for nuclear matter properties by employing the CC method on a momentum-space
lattice~\cite{hagen2013b} with a Bayesian machine-learning error model 
to quantify the uncertainties from the EFT
truncation~\cite{drischler2020} and the CC method (see
Methods and Extended Data Figure~\ref{fig:EOS_with_error} for details).
The observables we compute are the saturation density $\rho_0$, the energy per nucleon of
symmetric nuclear matter $E_0/A$, its compressibility $K$, the symmetry energy $S$ (i.e. the difference between the energy per nucleon of neutron matter and symmetric nuclear matter), and its slope $L$.
The posterior predictive distributions for these observables are shown in
Figure~\ref{fig:ppd_FW}a. These distributions include samples from the relevant method and model
error terms. Overall, we reveal relevant correlations among
observables, previously indicated in mean-field models, and find good agreement with empirical
bounds~\cite{lattimer2013}.
The last row shows the resulting correlations
with \rskinPb{} in our \emph{ab initio} framework.
In particular, we find essentially the same correlation between
\rskinPb{} and $L$ as observed in mean-field models (See Extended Data
Figure~\ref{fig:1S0_and_L_vs_Rskin}b).

\section{Discussion}
%
The predicted range of the \Pb{} neutron skin thickness (see Extended Data Table~\ref{tab:ppd})
is consistent with several extractions~\cite{ray1979,tarbert2013,roca-maza2015}, each of
which involves some model-dependence, and in mild tension
(approximately
$1.5 \sigma$) with the recent PREX
result~\cite{adhikari2021}.
{\it Ab initio} computations yield a thin skin and a narrow range  because the isovector physics is constrained by scattering data~\cite{pethick1995,tews2016,drischler2019}.
A thin skin was also predicted in \Ca{}~\cite{hagen2015}.

We find that both $\rskinPb{}=0.14-0.20$~fm and the slope parameter $L=37-66$~MeV are strongly correlated with scattering in the $^{1}S_0$ partial wave for laboratory energies around 50~MeV (the strongest two-neutron channel allowed by the Pauli principle, with the energy naively corresponding to the Fermi energy of neutron matter at $0.8 \rho_0$), see Extended Data Figure~\ref{fig:1S0_and_L_vs_Rskin}a.
It is possible, analogous to findings in mean-field theory~\cite{brown2000,todd-rutel2006}, to increase $L$ beyond
the range predicted in this work by tuning
a contact in the $^{1}S_0$ partial wave and simultaneously
readjusting the three-body contact to maintain realistic nuclear  saturation.
But this large slope $L$ and increased $R_{\rm skin}$ come at the cost of degraded $^{1}S_0$ scattering phase 
shifts, well beyond the expected corrections from higher-order terms (see Extended Data Figure~\ref{fig:parameter_vs_L}).
The large range of $L$ and $R_{\rm skin}$ obtained in mean-field theory is a consequence of scattering data not being incorporated.
It will be important to confront our predictions with more precise experimental measurements~\cite{CREX2022,mrexproposal}.
If the tension between scattering data and neutron skins persists, it will represent a serious challenge to our {\it ab initio} description of nuclear physics.

Our work demonstrates that {\it ab initio} approaches using nuclear forces from chiral EFT can consistently describe data from nucleon-nucleon scattering, few-body systems, and heavy nuclei within the estimated theoretical uncertainties.
Information contained in nucleon-nucleon scattering significantly constrains the properties of neutron matter; this same information constrains neutron skins, which provide a non-trivial empirical check on the reliability of {\it ab initio} predictions for the neutron matter equation of state.
Moving forward, it will be important to extend these calculations to higher orders in the effective field theory, both to further validate the error model and to improve precision,
and to push the cutoff to higher values to confirm regulator independence.
The framework presented in this work will enable predictions with quantified uncertainties across the nuclear chart, advancing toward the goal of a single unified framework for describing low energy nuclear physics.

\bibliographystyle{naturemag}
\bibliography{master,master_stats}

\titleformat{\subsection}[runin]{\small\raggedright\bfseries}{\arabic{section}.}{1em}{}

\clearpage

\section{\textcolor{blue}{METHODS}}


\setcounter{figure}{0}
\setcounter{table}{0}
\renewcommand{\figurename}{Extended Data Figure}
\renewcommand{\tablename}{Extended Data Table}
\renewcommand{\thefigure}{\arabic{figure}}
\renewcommand{\thetable}{\arabic{table}}
\renewcommand{\theHfigure}{Extended Data Figure\ \thefigure}
\renewcommand{\theHtable}{Extended Data Table \thetable}

\begin{figure*}[p]
    \centering
    \includegraphics[width=0.9\linewidth]{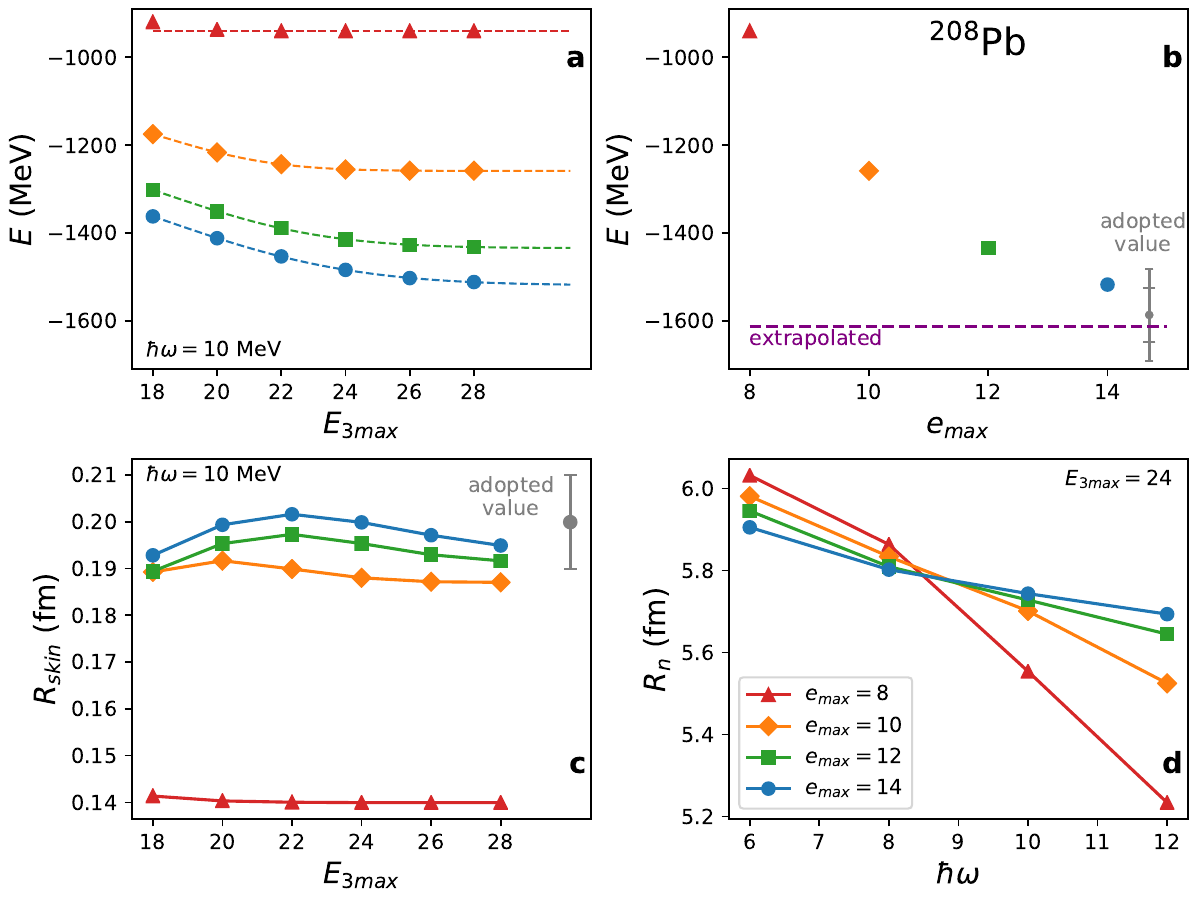}
    \caption[Convergence of energy and radius observables of \Pb{} with the ${e_{\rm max}}$ and ${E_{\rm 3max}}$ truncations.]{\textbf{Convergence of energy and radius observables of \Pbbold{} with the $\boldsymbol{e_{\rm max}}$ and $\boldsymbol{E_{\rm 3max}}$ truncations.} \textbf{a}, Ground state energy as a function of $E_{\rm 3max}$. The dashed lines indicate a Gaussian fit. \textbf{b}, Ground state energy (extrapolated in $E_{\rm 3max}$ as a function of $e_{\rm max}$. The smaller error bar on the adopted value indicate the error due to model space extrapolation, and the larger error bar also includes the method uncertainty. \textbf{c}, Neutron skin as a function of $E_{\rm 3max}$. \textbf{d}, Neutron radius as a function of oscillator basis frequency $\hbar\omega$ for a series of $e_{\rm max}$ cuts. }
    \label{fig:PbConvergence}
\end{figure*}

\textbf{\textcolor{blue}{Hamiltonian and model space.}}
 The many-body approaches used in this work [CC, IMSRG, and many-body perturbation theory (MBPT)] start from the intrinsic Hamiltonian
\begin{equation}
H = T_{\mathrm{kin}} - T_{\mathrm{CoM}} + V_\mathrm{NN} + V_\mathrm{3N}.
\end{equation}
Here $T_{\mathrm{kin}}$ is the kinetic energy, $T_{\mathrm{CoM}}$ is
the kinetic energy of the center of mass, $V_\mathrm{NN}$ is the
nucleon-nucleon, and $V_\mathrm{3N}$ is the three-nucleon
interaction. In order to facilitate the convergence of heavy nuclei,
the interactions employed in this work used a non-local regulator with
a cutoff $\Lambda = 394$~MeV$/c$. Specifically, the $V_\mathrm{NN}$ regulator is
$f(p)=\exp(p^2/\Lambda^2)^{n}$ and the $V_\mathrm{3N}$ regulator is
$f(p,q)=\exp[-(p^2+3q^2/4)/\Lambda^2]^{n}$  with $n=4$.
Results should be independent of this choice, up to higher-order
corrections, provided renormalization-group invariance of the EFT.
However, increasing the momentum scale of the cutoff leads to harder
interactions, considerably enlarging the required computational
effort.
We represent the 34 non-implausible interactions that resulted from the history-matching analysis in the Hartree-Fock basis in a model-space of up to 15 major harmonic oscillator shells ($e= 2n+l \le e_{\mathrm{max}}=14$ where $n$ and $l$ denote the radial and orbital angular momentum quantum numbers, respectively) 
with oscillator frequency $\hbar\omega = 10$~MeV.
Due to storage limitations, the three-nucleon force had an additional energy cut given by $e_1+e_2+e_3 \le E_{3\mathrm{max}} = 28$.
After obtaining the Hartree-Fock basis for each of the 34 non-implausible interactions, we capture 3N force effects via the normal-ordered two-body approximation before proceeding with the CC, IMSRG and MBPT calculations~\citemeth{hagen2007a, roth2012}.
The convergence behaviour in $e_{\rm max}$ and $E_{\rm 3max}$ is
illustrated in Extended Data Figure~\ref{fig:PbConvergence}.
In that figure, we use an interaction with a high likelihood that generates a large correlation energy. Thus, its convergence behaviour represents the worst case among the 34 non-implausible interactions.
The model-space converged results are investigated with $E_{\rm 3max} \to 3e_{\rm max}$ and $e_{\rm max} \to \infty$ extrapolations. The functions $E_{\rm gs} \approx c_{0} e^{-[(E_{\rm 3max}-c_{1})/c_{2}]^{2}}+E_{\rm gs}(E_{\rm 3max}=\infty)$ and $E_{\rm gs} \approx d_{0}e^{-d_{1} L_{\rm IR}} + E_{\rm gs}(e_{\rm max}=\infty)$ with $L_{\rm IR}=\sqrt{2(e_{\rm max}+7/2)b}$ ($b$ is the harmonic-oscillator length; $c_{i}$s and $d_{i}$s are the fitting parameters) are used as the asymptotic forms for $E_{\rm 3max}$~\cite{miyagi2021} and $e_{\rm max}$~\citemeth{furnstahl2014,furnstahl2014b}, respectively.
Through the extrapolations, the ground-state energies computed with $e_{\rm max}=14$ and $E_{\rm 3max}=28$ is shifted by $-75\pm 60$~MeV.
Likewise, the extrapolations of proton and neutron radii with the functional form given in Refs.~\cite{miyagi2021}$^,$\citemeth{furnstahl2014,furnstahl2014b} yield a small $+0.005\pm0.010$~fm shift of the neutron skin thickness.

\textbf{\textcolor{blue}{In-medium similarity renormalization group  calculations.}}
The IMSRG calculations~\cite{hergert2016}$^,$\citemeth{tsukiyama2011} 
were performed at the IMSRG(2) level, using the Magnus formulation~\citemeth{Morris2015}.
Operators for the point-proton and point-neutron radii, form factors, 
and the electric dipole operator were consistently transformed.
The dipole polarizablility $\alpha_D$ was computed using the equations-of-motion method truncated at the 2-particle-2-hole level (the EOM-IMSRG(2,2) approximation~\citemeth{parzuchowski2017}) and the Lanczos continued fraction method~\citemeth{miorelli2016}.
We compute the weak and charge form factors using the parameterization presented in Ref.~\citemeth{reinhard2013}, though
the form given in Ref.~\citemeth{hoferichter2020} yields nearly identical results.
%

\textbf{\textcolor{blue}{Many-body perturbation theory calculations.}}
MBPT theory calculations for $^{208}$Pb were performed in the Hartree-Fock basis to third order for the energies, and to second order for radii.

\textbf{\textcolor{blue}{Coupled-cluster calculations.}}
The CC calculations of $^{208}$Pb were truncated at the singles-and-doubles
excitation level, known as the CCSD approximation~\cite{kuemmel1978,bartlett2007,hagen2014}.
We estimated the contribution from triples excitations to the ground-state energy of
$^{208}$Pb as 10\% of the CCSD correlation energy (which is a reliable estimate for
closed-shell systems~\cite{bartlett2007}).

\begin{figure*}[p]
\includegraphics[width=0.975\linewidth]{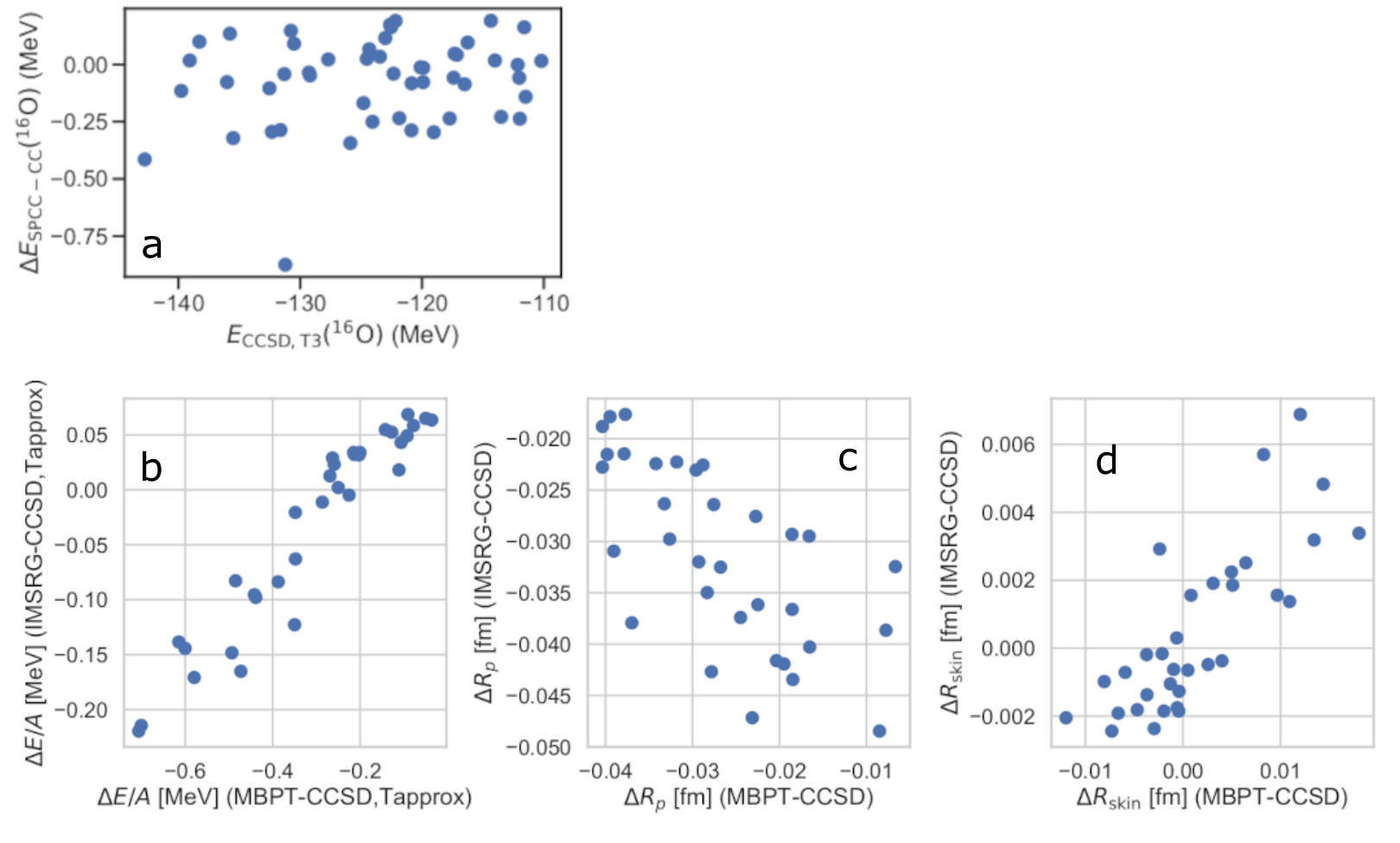}
\caption{ \textbf{Precision of sub-space coupled-cluster emulator and many-body solvers.}
  \textbf{a}, Cross-validation of the SP-CCSDT-3 emulator for the ground-state
  energy of $^{16}$O. Results from full computations using CCSDT-3 are
  compared with emulator predictions for 50 samples from the 17
  dimensional space of LECs. The standard deviation for the residuals
  $\Delta E_\mathrm{SPCC-CC}$ is {0.19}~MeV.
  \textbf{b,c,d}, Differences between IMSRG and CC results versus differences
    between MBPT and CC results for the ground-state energy per
    nucleon $\Delta E/A$ (panel b),
    the point-proton radius $\Delta R_\mathrm{p}$ (panel c), and the
    neutron-skin $\Delta R_\mathrm{skin}$ (panel d) of $^{208}$Pb using the 34 non-implausible
    interactions obtained from history matching (see text
    for more details). The CC results for the ground-state energy
    include approximate triples corrections.
    \label{fig:spcc_method_prec}
    }
  \end{figure*}

Extended Data Figure~\ref{fig:spcc_method_prec} compares the different many-body
approaches used in this work, i.e. CC, IMSRG, MBPT, and allows us to
estimate the uncertainties related to our many-body approach in
computing the ground-state observables for $^{208}$Pb. The point
proton and neutron radii are computed as ground-state expectation
values (see e.g. Ref.~\cite{hagen2015} for details). For $^{48}$Ca we
used a Hartree-Fock basis consisting of 15 major oscillator shells
with an oscillator spacing of $\hbar\omega = 16$~MeV, while for  3N
forces we used $E_{\mathrm{3max}} = 16$, which is sufficiently large to
obtain converged results in this mass region. Here we computed the
ground-state energy using the $\Lambda$-CCSD(T)
approximation~\citemeth{taube2008} which include perturbative triples
corrections. The $2^+$ excited state in $^{48}$Ca was computed using
the equation-of-motion CCSD approach~\citemeth{stanton1993}, and 
we estimated a $-1$~MeV shift from triples excitations based on EOM-CCSD(T)
calculations of $^{48}$Ca and $^{78}$Ni using similar
interactions~\citemeth{hagen2016b}.

For the history-matching analysis we used an emulator for the $^{16}$O
ground-state energy and charge radius that was constructed using the
recently developed sub-space coupled-cluster
method~\cite{ekstrom2019}. For higher precision in the emulator we
went beyond the SP-CCSD approximation used in Ref.~\cite{ekstrom2019}
and included leading-order triples excitations via the CCSDT-3
method~\citemeth{noga1987}. The CCSDT-3 ground-state training vectors
for $^{16}$O were obtained starting from the Hartree-Fock basis of the
recently developed chiral interaction
$\Delta$NNLO$_{\mathrm{GO}}$(394) of Ref.~\citemeth{jiang2020} in a
model-space consisting of 11 major harmonic oscillator shells with the
oscillator frequency $\hbar\omega = 16$~MeV, and $E_{\mathrm{3max}} =
14$. The emulator used in the history matching was
constructed by selecting 68 different training points in the
17-dimensional space of LECs using a space-filling Latin hypercube
design in a 10\% variation around the
$\Delta$NNLO$_{\mathrm{GO}}$(394) LECs. At each training point we then
performed a CCSDT-3 calculation in order to obtain the training
vectors for which we then construct the sub-space projected norm and
Hamiltonian matrices. Once the SP-CCSDT-3 matrices are constructed we
may obtain the ground-state energy and charge radii for any target
values of the LECs by diagonalizing a 68 by 68 generalized eigenvalue
problem (see Ref.~\cite{ekstrom2019} for more details). We checked the
accuracy of the emulator by cross-validation against full-space
CCSDT-3 calculations as demonstrated in Extended Data Figure~\ref{fig:spcc_method_prec}a and
found a relative error that was smaller than 0.2\%.

The nuclear matter equation of state and saturation properties are
computed with the CCD(T) approximation which includes doubles
excitations and perturbative triples corrections. The three-nucleon
forces are considered beyond the normal-ordered two-body approximation
by including the residual three-nucleon force contribution in the
triples. The calculations are performed on a cubic lattice in momentum
space with periodic boundary conditions. The model space is
constructed with
$(2n_{\rm max}+1)^3$ momentum points, and we use $n_{\rm max} = 4 (3)$
for
pure neutron matter (symmetric nuclear matter)
and obtain converged results. We perform calculations for systems of 66 neutrons (132 nucleons)
for pure neutron matter (symmetric nuclear matter) since results obtained with those particle numbers exhibit small finite
size effects~\cite{hagen2013b}. 

\textbf{\textcolor{blue}{Iterative history matching.}}
In this work we use an iterative approach 
known as history matching~\cite{Vernon:2010,Vernon:2014} in which the model, 
solved at different fidelities, is
confronted with experimental data $z$ using
Eq.~\eqref{eq:modelreality}.
Obviously, we do not know the exact values of the errors in
Eq.~\eqref{eq:modelreality}, 
hence we represent them as random variables and specify 
reasonable forms for their statistical distributions, in alignment
with the Bayesian paradigm.

For many-body systems we employ quantified method and ($A=16$) emulator errors as discussed
above and summarized in Extended Data Table~\ref{tab:error_assignments}.
For $A \leq 4$ nuclei we use the no-core shell model in Jacobi
coordinates~\citemeth{kamuntavicius2000} and eigenvector continuation emulators~\cite{Konig:2019adq}. The associated
method and emulator errors are very small.
Probabilistic attributes of the model discrepancy
terms are assigned based on the expected EFT convergence
pattern~\citemeth{Wesolowski:2015fqa, Melendez:2017phj}. For the
history-matching observables considered here we use point estimates of model errors
from Ref.~\citemeth{ekstrom2018}.

The aim of history matching is to estimate the set $\mathcal{Q}(z)$ of parameterizations
$\theta$, for which the evaluation of a model $M(\theta)$ yields an
acceptable---or at least not implausible---match to a set of
observations $z$. History matching has been employed in various
studies involving complex computer models~\citemeth{Craig96_Pressure,Craig97_Pressure,Vernon:2010-2,Andrianakis:2015}
ranging, e.g., from effects of climate
modeling~\citemeth{Williamson:2013,Edwards:2019tp} to systems
biology~\citemeth{Vernon:2018}.

We introduce the individual implausibility measure
\begin{equation}
  I_i^2(\theta) = \frac{|{M}_i(\theta) -
    z_i|^2}{\mathrm{Var} \left( {M}_i(\theta) - z_i \right)},
  \label{eq:IMi}
\end{equation}
which is a function over the input parameter space and quantifies the
(mis-)match between our (emulated) model output ${M}_i(\theta)$ and the observation
$z_i$ for an observable in the target set $\mathcal{Z}$.
We mainly employ a maximum implausibility measure as the restricting quantity. Specifically, we
consider a particular value for $\theta$ as implausible if
\begin{equation}
  I_M(\theta) \equiv \max_{z_i \in \mathcal{Z}} I_i(\theta) > c_I,
  \label{eq:IM}
\end{equation}
with $c_I \equiv 3.0$ appealing to Pukelheim's three-sigma
rule~\citemeth{Pukelsheim:1994}.
In accordance with the assumptions leading to
Eq.~\eqref{eq:modelreality}, the variance in the denominator of
Eq.~\eqref{eq:IMi} is a sum of independent squared
errors. Generalizations of these assumptions are straightforward if
additional information on error covariances or possible inaccuracies
in our error model would become available.

\begin{figure*}[p]
  \includegraphics[width=0.975\linewidth]{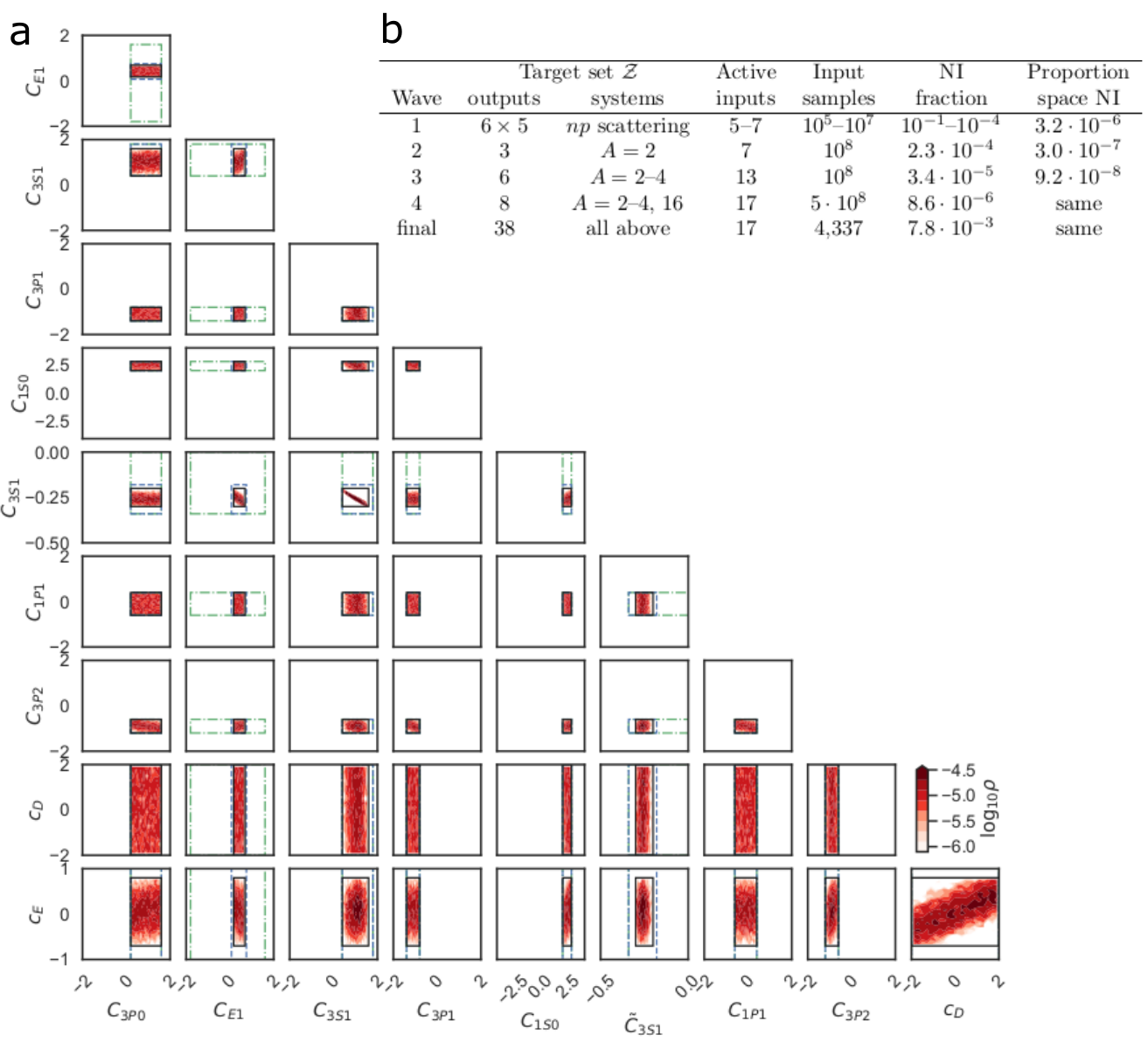}
  \caption{ \textbf{History-matching waves.}
    \textbf{a}, The initial parameter domain used at the start of
    history-matching wave~1 is represented by the axes limits for all
    panels. This domain is iteratively reduced and the input volumes
    of waves 2, 3, and 4 are indicated by green/dash-dotted,
    blue/dashed, black/solid rectangles. The logarithm of the optical depths $\log_{10}\rho$ (indicating
    the density of non-implausible samples in the final wave) are
    shown in red with darker regions corresponding to a denser
    distribution of non-implausible samples. 
    \textbf{b},
    Four waves of history matching were used in this work plus a fifth one to 
   refine the final set of non-implausible samples.
   The neutron-proton scattering targets correspond to  phase
 shifts at six energies ($T_\mathrm{lab} = 1, 5, 25, 50,
 100, 200$ MeV) per partial wave: ${}^1S_0$, ${}^3S_1$, ${}^1P_1$, ${}^3P_0$,
 ${}^3P_1$, ${}^3P_2$. The $A=2$ observables are $E(^{2}\mathrm{H})$,  $R_\mathrm{p}(^{2}\mathrm{H})$,
        $Q(^{2}\mathrm{H})$, while $A=3,4$ are $E({}^{3}\mathrm{H})$, $E({}^{4}\mathrm{He})$,
        $R_\mathrm{p}({}^{4}\mathrm{He})$. Finally, $A=16$ targets are $E({}^{16}\mathrm{O})$,
        $R_\mathrm{p}({}^{16}\mathrm{O})$. The number of active input
        parameters is indicated in the fourth column. The number of inputs
        sets being explored, and the fraction of non-implausible
        samples that survive the imposed implausibility
        cutoff(s) are shown in the fifth and sixth columns, respectively.
        Finally, the
        proportion of the parameter space deemed non implausible is listed in the
        last column. Note that no additional reduction of the non-implausible domain is
        achieved in the fourth and final waves, in which
        ${}^{16}\mathrm{O}$ observables are included, but that parameter correlations are enhanced.
    \label{fig:hm_waves}
    }
\end{figure*}

\begin{figure}[p]
\includegraphics[width=1.0\linewidth]{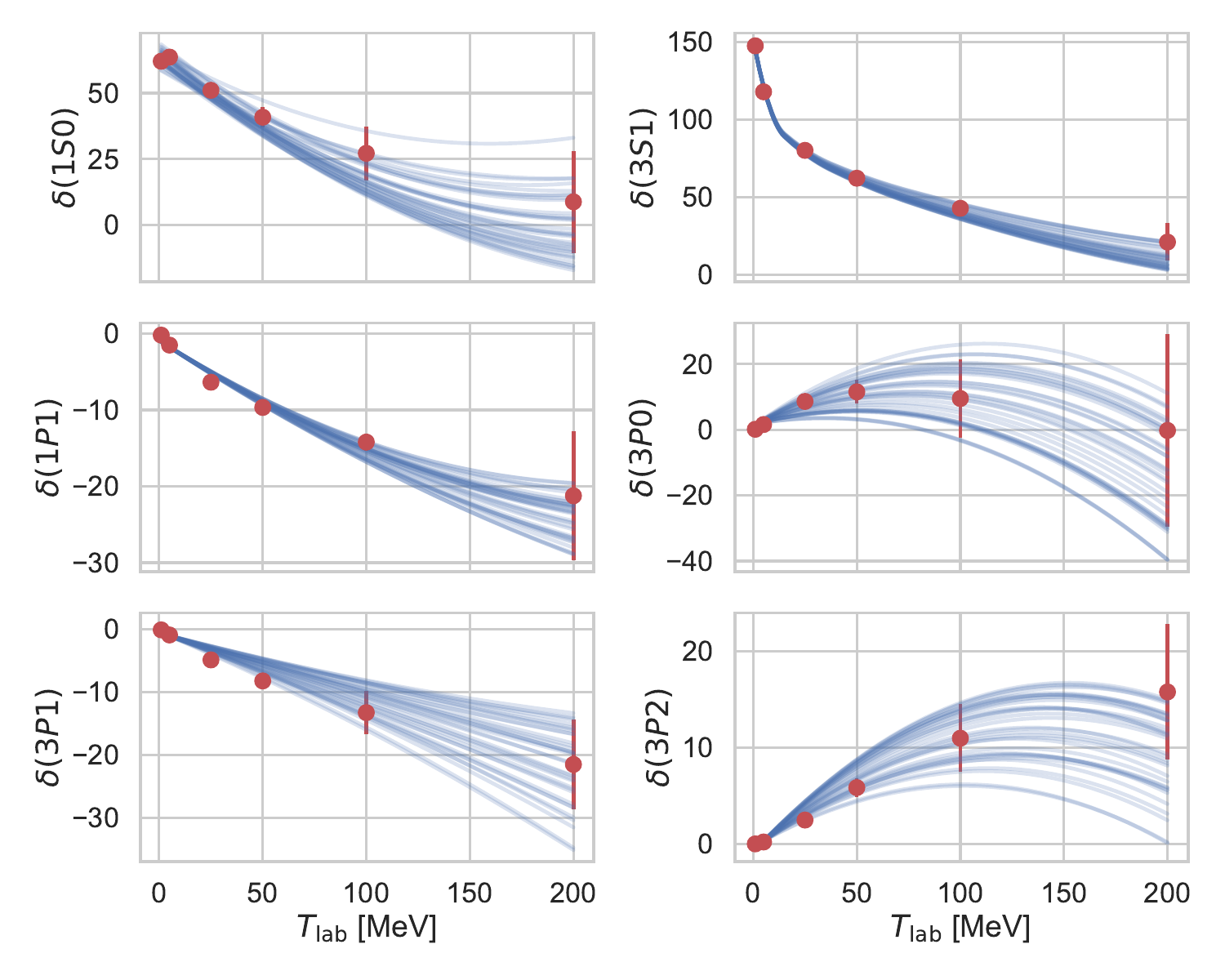}
\caption[Neutron-proton scattering phase shifts.]
{ \textbf{Neutron-proton scattering phase shifts.}
   34 interaction samples survive the final implausibility cutoff with
   respect to neutron-proton phase shifts $\delta$ in $S$ and $P$ waves up to
   200~MeV. The red circles are from the Granada phase shift
   analysis\citemeth{navarro2013}, while the $2\sigma$ error bars are
   dominated by the estimated EFT truncation errors\citemeth{ekstrom2018}. 
    \label{fig:NI_phase_shifts}
    }
\end{figure}

\begin{figure*}[p]
    \begin{subfigure}[b]{0.48\textwidth}
      \textbf{a}\hspace*{0.8\textwidth}\mbox{}\\
    \centering
    \includegraphics[width=\columnwidth]{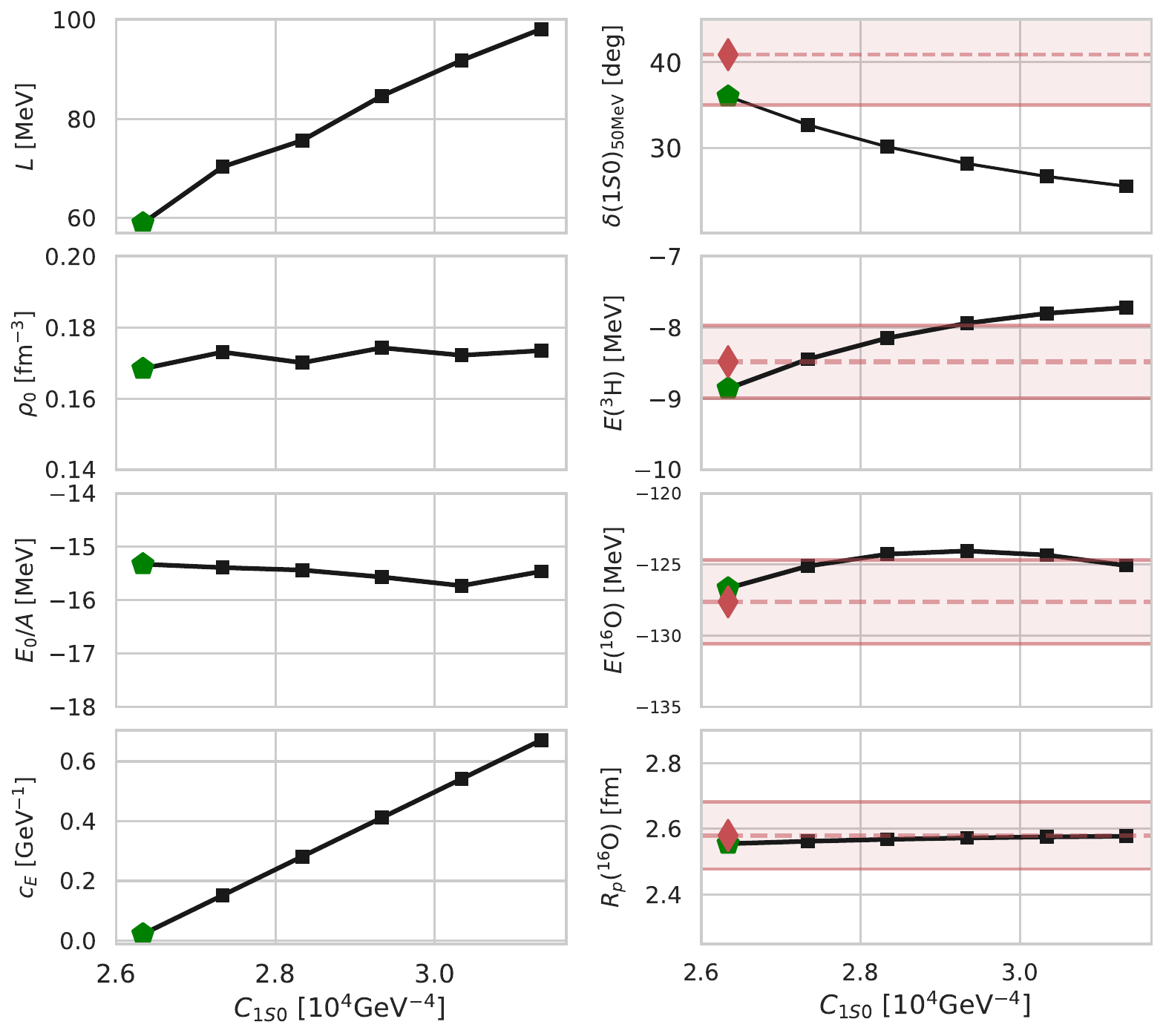}
  \end{subfigure}
  \hfill
  \begin{subfigure}[b]{0.48\textwidth}
    \textbf{b}\hspace*{0.8\textwidth}\mbox{}\\
    \centering
    \includegraphics[width=\columnwidth]{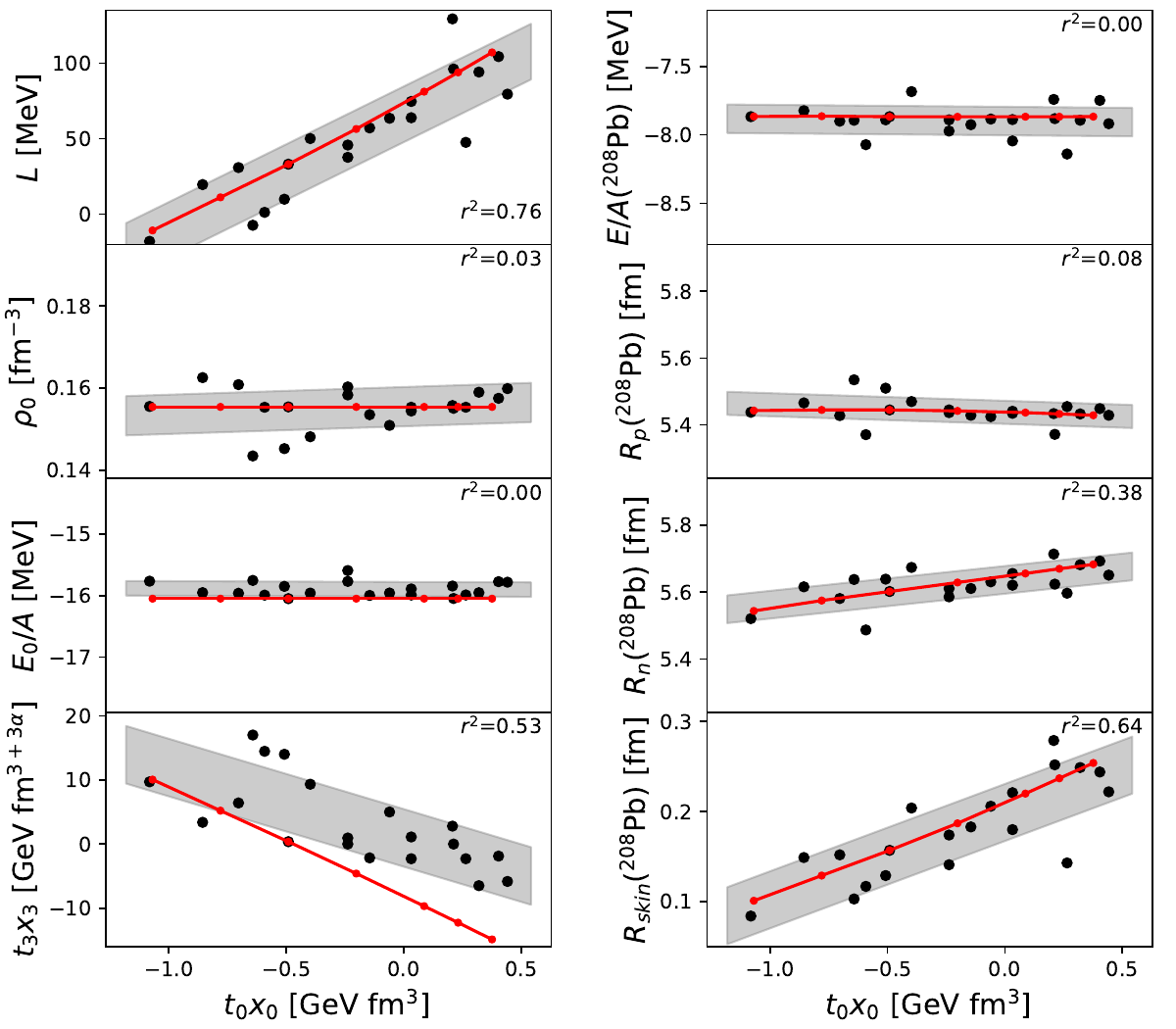}
  \end{subfigure}
  
  \caption[Parameter sensitivity]{\textbf{Parameter sensitivities in
      \textit{ab initio} models and Skyrme parametrizations.}
      \textbf{a}, Tuning the $C_{1S0}$ LEC in our \textit{ab initio} model to adjust the symmetry energy slope
      parameter $L$ while compensating with the three-nucleon contact $c_E$ to
      maintain the saturation density $\rho_0$
      and energy per nucleon $E_0/A$ of symmetric nuclear matter.
      The green pentagons correspond to results with one of the 34 interaction samples while the black squares indicate the results after tuning the $C_{1S0}$ and $c_E$ of that interaction.
      The right column shows the scattering phase shift $\delta$ in the $^{1}S_0$
      channel at 50~MeV, the ground-state energies in $^3\mathrm{H}$ and \Ox{} and the point-proton radius $R_\mathrm{p}$ in \Ox. The red diamonds and the dashed lines indicate the experimental values of target observables and the red bands indicate the corresponding $c_I=3$ non-implausible regions, see Eq.~\eqref{eq:IM}, Extended Data Table~\ref{tab:error_assignments} and Extended Data Figure~\ref{fig:NI_phase_shifts}.
      \textbf{b}, Illustration of the freedom in Skyrme
        parametrizations to
      adjust $L$ while preserving $\rho_0$ and $E_0/A$. 
      The parameters $x_0,t_0,x_3,t_3$ correspond to the functional form given in e.g.~\citemeth{brown1988}. The black circles correspond to different parameter sets, while the red line indicates the result of starting with the SKX interaction and modifying the $x_0,x_3$ parameters while maintaining the binding energy per  nucleon $E/A$ of $^{208}$Pb. The right column also shows the \Pb{} point-proton and point-neutron radii ($R_\mathrm{p}$ and $R_\mathrm{n}$, respectively) and neutron skin thickness $R_\mathrm{skin}$ for different parametrisations. The gray bands indicate a linear fit to the black points with $r^2$ the coefficient of determination. Skyrme parameter
      sets included are SKX, SKXCSB~\cite{brown1998}, SKI,
      SKII~\citemeth{vautherin1972}, SKIII-VI~\citemeth{beiner1975},
      SKa, SKb~\citemeth{kohler1976}, SKI2,
    SKI5~\citemeth{reinhard1995}, SKT4, SKT6~\citemeth{tondeur1984},
      SKP~\citemeth{dobaczewski1984}, SGI,
      SGII~\citemeth{vangiai1981}, MSKA~\citemeth{sharma1995},
      SKO~\citemeth{reinhard1999}, SKM$^*$~\citemeth{bartel1982}. 
      \label{fig:parameter_vs_L}
      }
\end{figure*}

An important strength of the history matching is that we can proceed
iteratively, excluding regions of input space by imposing cutoffs on
implausibility measures that can include \textit{additional}
observables $z_i$ and corresponding model outputs $M_i$ with possibly
refined emulators as the parameter volume is reduced. The history matching process is designed to be independent of the order in which observables are included, as is discussed in~\citemeth{Vernon:2010-2}. This is an important feature as it allows for efficient choices regarding such orderings.
The iterative history matching proceeds in waves according to a straightforward
strategy that can be summarized as follows:
\begin{enumerate}
  \item At wave $j$: Evaluate a set of model runs over the
    current NI volume $\mathcal{Q}_j$ using a space-filling design of
    sample values for the parameter inputs $\theta$. Choose a
    rejection strategy based on implausibility measures for a set
    $\mathcal{Z}_j$ of informative observables.
  \item Construct or refine emulators for the model
    predictions across $\mathcal{Q}_j$.
  \item The implausibility measures are then calculated over
    $\mathcal{Q}_j$ using the emulators, and implausibility
    cutoffs are imposed. This defines a new, smaller non-implausible
    volume $\mathcal{Q}_{j+1}$ which should satisfy $\mathcal{Q}_{j+1}
    \subset \mathcal{Q}_{j}$.
      \item Unless (a) computational resources are
        exhausted, or (b) all considered points in the parameter space
        are deemed implausible, we may include additional informative observables in the
        considered set $\mathcal{Z}_{j+1}$, and return to step 1.
    
    \item If 4(a) is true we generate a number of acceptable runs from the
        final non-implausible volume $\mathcal{Q}_\mathrm{final}$, sampled according to scientific need.
\end{enumerate}
The \textit{ab initio} model for the observables we consider comprises
at most 17 parameters; four subleading pion-nucleon couplings, 11
nucleon-nucleon contact couplings, and two short-ranged three-nucleon
couplings. To identify a set of non-implausible parameter samples we performed
iterative history matching in four waves using observables and implausibility
measures as summarized in Extended Data
Figure~\ref{fig:hm_waves}b. For each wave we employ a sufficiently
dense Latin hypercube set of several million candidate parameter
samples. For the model evaluations we utilized fast computations of
neutron-proton scattering phase shifts and efficient emulators
for the few- and many-body history-matching observables. See Extended Data
Table~\ref{tab:error_assignments} and Extended Data
Figure~\ref{fig:NI_phase_shifts} for the list of history-matching
observables and information on the errors that enter the
implausibility measure~\eqref{eq:IMi}. The input volume for wave~1
incorporates the naturalness expectation for LECs, but still includes
large ranges for the relevant parameters as indicated by the panel
ranges in Extended Data Figure~\ref{fig:hm_waves}a.
In all four waves the input volume for $c_{1,2,3,4}$ is a
four-dimensional hypercube mapped onto the
multivariate Gaussian probability density function (PDF) resulting from a 
Roy-Steiner analysis of pion-nucleon
scattering data~\citemeth{Siemens:2017}.
In wave~1 and wave~2 we
sampled all relevant parameter directions for the set of included
two-nucleon observables. In wave 3, the $^3$H and $^4$He
observables were added such that the three-nucleon force parameters $c_D$ and $c_E$
can also be constrained. Since these observables are known to be rather insensitive
to the four model parameters acting solely in $P-$waves, we
ignored this subset of the inputs and compensated by
slightly enlarging the corresponding method errors.
This is a well known emulation procedure called inactive parameter
identification~\cite{Vernon:2010}. 
For wave~4 we considered all 17 model parameters and
added the ground-state energy and radius of $^{16}$O to the set $\mathcal{Z}_4$ and
emulated the model outputs for $5\times 10^{8}$ parameter
samples. By including oxygen data we explore the modeling capabilities of
our \textit{ab initio} approach. Extended Data
Figure~\ref{fig:hm_waves}a summarizes the sequential non-implausible
volume reduction, wave-by-wave, and indicates the set of 4,337
non-implausible samples after the fourth wave. We note that the use of
history matching would in principle allow a detailed study of the
information content of various observables in heavy-mass nuclei. Such
an analysis, however, requires an extensive set of reliable emulators and is beyond the scope of the present work.
The volume reduction is determined by the maximum implausibility cutoff~\eqref{eq:IM} with additional confirmation from the
optical depths (which indicate the density of non-implausible samples; see Eqs.~(25) and (26) in
Ref.~\citemeth{Vernon:2018}). 
The non-implausible samples summarise the parameter region of interest, and can
directly aid insight regarding interdependencies between parameters
induced by the match to observed data. This region is also where we
would expect the posterior distribution to reside and we note that our
history-matching procedure has allowed us to reduce its size by more than seven orders of magnitude
compared to the prior volume (see Extended Data Figure~\ref{fig:hm_waves}b). 

As a final step, we confront the set of non-implausible samples from wave~4 with
neutron-proton scattering phase shifts such that our final set of non-implausible samples
has been matched with all history-matching observables. For this final
implausibility check we employ a slightly less strict cutoff and
allow the first, second and third maxima of $I_i(\theta)$ (for $z_i \in \mathcal{Z}_\mathrm{final}$) to be 5.0, 4.0, and 3.0,
respectively, accommodating the more extreme maxima we may anticipate
when considering a significantly larger number of observables. The end
result is a set of 34 non-implausible samples that we use for 
predicting \Ca{} and \Pb{} observables, as well as the equation of
state of both symmetric nuclear matter and pure neutron matter.

\textbf{\textcolor{blue}{Posterior predictive distributions.}}
The 34 non-implausible samples from the final history matching wave are used to compute energies,
radii of proton and neutron distributions, and electric dipole polarizabilities ($\alpha_D$)for \Ca{} and \Pb{}. They are
also used to compute the electric and weak charge
form factors for the same nuclei at a relevant momentum transfer, and the
energy per particle of infinite nuclear matter at various densities
to extract key properties of the nuclear equation of state (see below). These results are shown as blue circles in
Figure~\ref{fig:ppd_FW}. 

In order to make quantitative predictions, with a statistical interpretation, for \rskinPb{} and other
observables we use the same 34
parameter sets to extract representative samples from the
posterior PDF
$p(\theta|\mathcal{D}_\mathrm{cal})$. Bulk
properties (energies and charge radii)
of \Ca{} together with the structure-sensitive $2^+$
excited-state energy of \Ca{} are used to define the calibration data
set $\mathcal{D}_\mathrm{cal}$. The IMSRG and CC convergence studies
make it possible to quantify the method errors. These are summarized in Extended Data
Table~\ref{tab:error_assignments}. The
EFT truncation errors are quantified by adopting the EFT convergence model~\citemeth{furnstahl2015a,Melendez2019} for observable $y$
\begin{equation}
  y = y_\mathrm{ref} \left( \sum_{i=0}^k c_i Q^i + \sum_{i=k+1}^\infty c_i
    Q^i \right),
  \label{eq:EFTconvergence}
\end{equation}
with observable coefficients $c_i$ that are expected to be of natural
size, and the expansion parameter Q = 0.42
  following our Bayesian error model for nuclear matter at the
  relevant density (see below).
The first sum in the parenthesis is the model prediction $y_k(\theta)$ of
observable $y$ at truncation order $k$ in the chiral expansion. The second sum
than represents the model error as it includes the terms that are not explicitly included. We can
quantify the magnitude of these terms by learning about the
distribution for $c_i$ which we will assume is described by a single
normal distribution per observable type with zero mean and a variance parameter $\bar{c}^2$.
We employ the nuclear matter error analysis for the energy
per particle of symmetric nuclear matter (described below) to
provide the model error for $E/A$ in \Ca{} and \Pb{}. For radii and
electric dipole polarizabilities we employ the next-to leading order 
and next-to-next-to leading order
interactions of Ref.~\citemeth{jiang2020} and compute these observables at
both orders for various Ca, Ni, and Sn isotopes. The reference values
$y_\mathrm{ref}$ are set to $r_0 \cdot A^{1/3}$ for radii and to the
experimental value for $\alpha_D$. From this data we extract
$\bar{c}^2$ and perform the geometric sum of the second term in
Eq.~\eqref{eq:EFTconvergence}. The resulting standard deviations for
model errors are summarized in Extended Data
Table~\ref{tab:error_assignments}.

At this stage we can approximately extract samples from the parameter posterior $p(\theta
  | \mathcal{D}_\mathrm{cal})$ by employing the established method
  of sampling/importance
resampling~\cite{Smith:1992aa}$^,$\citemeth{bernardo2006bayesian}.
We assume a uniform prior probability for the
non-implausible samples and we introduce a
normally-distributed likelihood, $\mathcal{L}(\mathcal{D}_\mathrm{cal}
| \theta)$, assuming independent experimental, method,
and model errors.
The prior for $c_{1,2,3,4}$ is the
multivariate Gaussian resulting from a Roy-Steiner analysis of $\pi N$
scattering data~\citemeth{Siemens:2017}. 
Defining importance weights
\begin{equation}
  q_i = \mathcal{L}(\mathcal{D}_\mathrm{cal}
| \theta_i) / \sum_{j=1}^n \mathcal{L}(\mathcal{D}_\mathrm{cal} |
\theta_j),
\label{eq:importanceweights}
\end{equation}
we draw samples $\theta^*$ from the discrete distribution
$\{\theta_1, \ldots, \theta_n\}$ with probability mass $q_i$ on
$\theta_i$. These samples are then approximately distributed according
to the parameter posterior that we are
seeking~\cite{Smith:1992aa}$^,$\citemeth{bernardo2006bayesian}. 

\begin{figure}[p]
\includegraphics[width=1.0\linewidth]{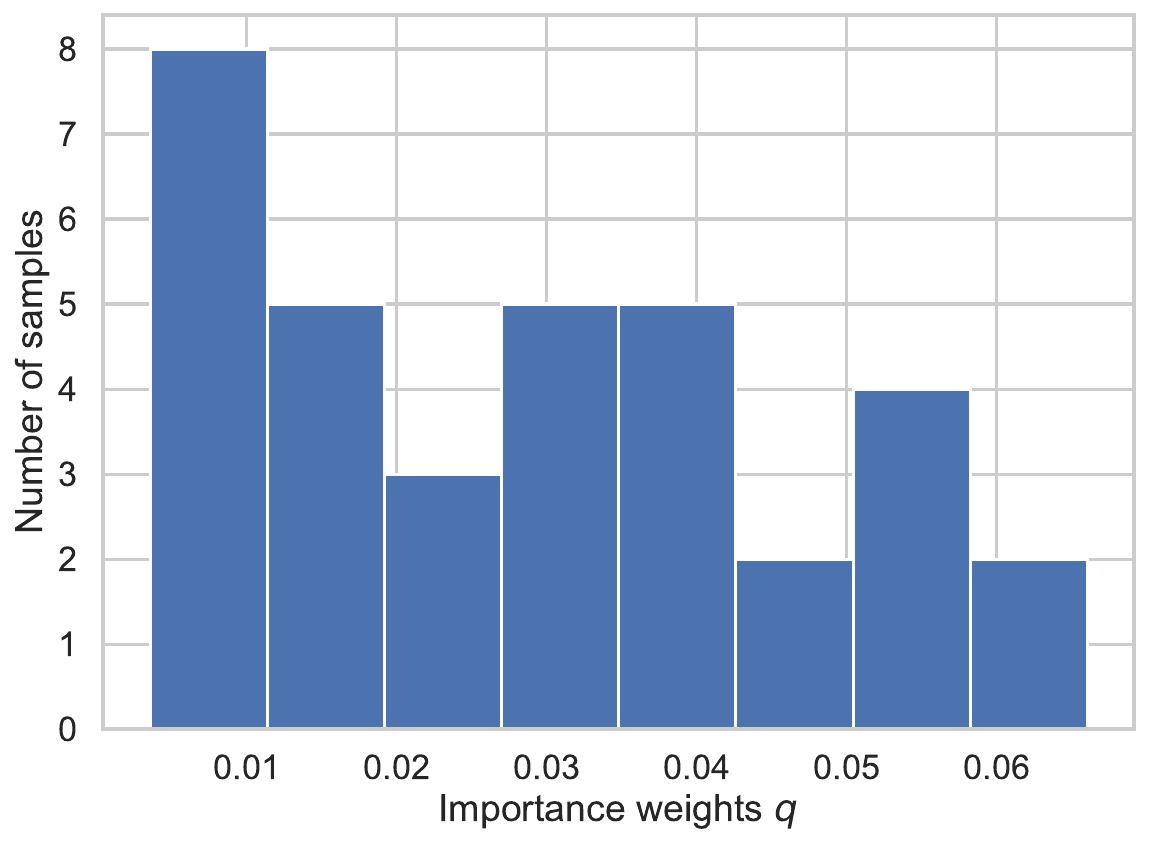}
\caption[Importance weights.]
{ \textbf{Importance weights.}
   Histogram of importance weights for the 34 non-implausible interaction 
   samples. These are obtained from likelihood calibration as
   defined in Eq.~\eqref{eq:importanceweights}. 
    \label{fig:importanceweights}
  }
\end{figure}

Although we are operating with a finite number of $n=34$ representative samples from the parameter PDF, it is reassuring that about half of them are within a factor two from the most probable one in terms of the importance weight, see Extended Data Figure~\ref{fig:importanceweights}. Consequently, our final predictions will not be dominated by a very small number of interactions. In addition, as we do not anticipate the parameter PDF to be of a particularly complex shape, based on the results of the history match, consideration of the various error structures in the analysis, and on the posterior predictive distributions (PPDs) shown in Figure~\ref{fig:ppd_FW}, and as we are mainly interested in examining such lower 1- or 2-dimensional PPDs, this sample size was deemed sufficient and the corresponding sampling error assumed subdominant. We use these samples to draw corresponding samples from 
\begin{equation}
   \text{PPD}_{\text{parametric}} = \{y_k(\theta) : \theta \sim p(\theta
  | \mathcal{D}_\mathrm{cal})\}.
  \label{eq:trunc_model_ppd}
\end{equation}
This PPD is the set of all model predictions computed over likely
values of the parameters, i.e., drawing from the posterior PDF for
$\theta$.
The full PPD is then defined, in analogy with
Eq.~\eqref{eq:trunc_model_ppd},  as the set evaluation of
$y$ which is the sum
\begin{equation}
  y = y_k + \epsilon_\mathrm{method} + \epsilon_\mathrm{model},
  \label{eq:fullstatmodel}
\end{equation}
where we assume method and model errors to be independent of the
parameters. In practice, we produce $10^4$ samples from this full PPD
for $y$ by resampling the 34 samples of the model
PPD~\eqref{eq:trunc_model_ppd} according to their importance weights, and adding
samples from the error terms in~\eqref{eq:fullstatmodel}. We perform
model checking by comparing this final PPD with the data used in the
iterative history-matching step, and in the likelihood calibration. In addition, we
find that our predictions for the measured electric dipole
polarizabilities of \Ca{} and \Pb{} as well as bulk properties of \Pb{} serve as a validation of the
reliability of our analysis and assigned errors.  See
Figure~\ref{fig:blobbogram} and Extended Data
Table~\ref{tab:error_assignments}.

In addition, we explored the sensitivity of our results to
modifications of the likelihood definition. Specifically, we used a
student-t distribution ($\nu=5$) to see the effects of allowing
heavier tails, and we introduced an error covariance matrix to study
the effect of possible correlations (with $\rho \approx 0.7$) between
the errors in binding energy and radius of \Ca{}. In the end, the differences in the extracted credibility
regions was $\sim 1\%$ and we therefore present only results obtained
with the uncorrelated, multivariate normal distribution.

Our final predictions for \rskinPb{}, \rskinCa{} and for nuclear
matter properties are presented in Figure~\ref{fig:ppd_FW} and Extended
Data Table~\ref{tab:ppd}. For these observables we use the Bayesian
machine learning error model described below to assign relevant
correlations between equation-of-state observables. For model errors in \rskinPb{} and $L$ we use a correlation
coefficient of $\rho=0.9$ as motivated by the strong
correlation between the observables computed with the 34
non-implausible samples. It should be noted that $S$, $L$, and $K$ are
computed at the specific saturation density of the corresponding
non-implausible interaction.

\textbf{\textcolor{blue}{Bayesian machine learning error model.}}
 Similar to Eq.~\eqref{eq:modelreality} the predicted nuclear matter
 observables can be written as:
\begin{equation}
y =  y_k (\rho)  + \varepsilon_k(\rho) + \varepsilon_\mathrm{method}(\rho).
\label{eq:predicted_observables}
\end{equation}
where $y_k (\rho)$ is the CC prediction using our EFT model truncated
at order $k$, $\varepsilon_k(\rho)$ is the EFT truncation (model)
error, and $\varepsilon_\mathrm{method}(\rho)$ 
is the CC method error. In this work we apply a Bayesian machine learning error
model~\cite{drischler2020} to quantify the density dependence of both method and truncation errors. The error
model is based on multitask Gaussian processes that learn
both the size and the correlations of the target errors from given
prior information. Following a physically-motivated Gaussian process (GP) model~\cite{drischler2020}, the
EFT truncation errors $\varepsilon_k$ at given density $\rho$ are
distributed as:
\begin{equation}
\varepsilon_k(\rho) ~ |~ \bar{c}^2,l,Q \sim {\rm GP}[0,\bar{c}^2R_{\varepsilon_k}(\rho,\rho';l)],
\label{eq:truncation_error_1}
\end{equation}
with
\begin{equation}
  R_{\varepsilon_k}(\rho,\rho';l) = y_{\rm{ref}}(\rho)y_{\rm{ref}}(\rho')
  \frac{[Q(\rho)Q(\rho')]^{k+1}}{1-Q(\rho)Q(\rho')}r(\rho,\rho';l).  
\label{eq:truncation_error_2}
\end{equation}
Here $k=3$ for the $\Delta$NNLO(394) EFT model used in this work, while $\bar{c}^2$, $l$ and
$Q$ are hyperparameters corresponding to the variance, the correlation
length, and the expansion parameter.
Finally, we choose the reference scale $y_{\rm{ref}}$ to be the EFT leading-order prediction. The mean of the Gaussian process is set to be zero
since the order-by-order truncation error can either be positive or
negative and the correlation function $r(\rho, \rho'; l)$ in~\eqref{eq:truncation_error_2} is
the Gaussian radial basis function.

We employ Bayesian inference to optimize the Gaussian process hyperparameters using order-by-order predictions of the equation of state for both pure neutron matter and symmetric nuclear matter with the $\Delta$-full interactions from
Ref.~\citemeth{ekstrom2018}. In this work, we find $\bar{c}_{\rm{PNM}}=1.00$ and $l_{\rm{PNM}}= 0.92~\rm{fm}^{-1}$
for pure neutron matter and $\bar{c}_{\rm{SNM}}=1.55$ and $l_{\rm{SNM}}= 0.48~\rm{fm}^{-1}$ for symmetric nuclear matter.

The above Gaussian processes only describe the correlated structure of
truncation errors for one type of nucleonic matter. In addition, the correlation between pure neutron matter and symmetric nuclear matter is
crucial for correctly assigning errors to observables that involve both $E/N$ and
$E/A$ (such as the symmetry energy $S$). For this purpose we use a multitask Gaussian process that
simultaneously describes truncation errors of pure neutron matter and
symmetric nuclear matter according to:
\begin{equation}
\begin{bmatrix}
\varepsilon_{k,\rm{PNM}}\\
\varepsilon_{k,\rm{SNM}}
\end{bmatrix} \sim {\rm GP} \left(
\begin{bmatrix}
0\\
0
\end{bmatrix},
\begin{bmatrix}
K_{11}& K_{12}\\
K_{21}& K_{22}
\end{bmatrix}  \right) ,
\label{eq:multitask_GP}
\end{equation}
where $K_{11}$ and $K_{22}$ are the covariance matrices generated from
the kernel function $\bar{c}^2R_{\varepsilon_k}(\rho,\rho';l)$ for pure neutron matter and
symmetric nuclear matter, respectively, while $K_{12}$($K_{21}$) is the cross-covariance as in
Ref.~\citemeth{Drischler2020b}.

Regarding the CC method error, different sources of uncertainty should be considered. 
The truncation of the cluster operator and the finite-size effect are the main ones and the total method error is then $\varepsilon_\mathrm{method}=\varepsilon_\mathrm{cc}+\varepsilon_\mathrm{fs}$.
Following the Bayesian error model we have a general expression for the method error:
\begin{equation}
\varepsilon_\mathrm{me}(\rho) ~ |~ \bar{c}_{\rm{me}}^2,l_{\rm{me}}, \sim {\rm GP}[0,\bar{c}_{\rm{me}}^2R_{\rm{me}}(\rho,\rho';l_{\rm{me}})],
\label{eq:method_error_1}
\end{equation}
with
\begin{equation}
R_{\rm{me}}(\rho,\rho';l_{\rm{me}}) = y_{\rm{me, ref}}(\rho)y_{\rm{me, ref}}(\rho') r(\rho,\rho';l_{\rm{me}}). 
\label{eq:method_error_2}
\end{equation}
Here the subscript ``$\rm{me}$" stands for either the cluster operator
truncation ``$\rm{cc}$" or finite-size effect ``$\rm{fs}$" method
error.
For the cluster operator truncation errors $\varepsilon_\mathrm{cc}$ the reference scale $y_{\rm{me,ref}}$ is taken to be the CCD(T) correlation energy. The Gaussian processes are then optimized with data from different
interactions by assuming that the energy difference between CCD and
CCD(T) can be used as an approximation of the cluster operator truncation error. The
correlation lengths learned from the training data are
$l_{\rm{me, PNM}}= 0.81~\rm{fm}^{-1}$ for pure neutron matter and $l_{\rm{me, SNM}}= 0.34~\rm{fm}^{-1}$ for
symmetric nuclear matter. 
Based on the convergence study we take $\pm10\%$ of the correlation energy as the 95\% credible interval which gives $\bar {c}_{\rm{me}} = 0.05$ for $\varepsilon_\mathrm{cc}$.
As for the finite-size effect $\varepsilon_\mathrm{fs}$, the reference scale is taken to be the CCD(T) ground-state energy.
Then following Ref.~\citemeth{hagen2013b}, we use $\pm0.5\%$ ($\pm4\%$) of the ground-state energy of 
the pure neutron matter (the symmetric nuclear matter) 
as a conservative estimation of the finite-size effect (95\% credible interval) when using periodic boundary conditions 
with 66 neutrons (132 nucleons) around the saturation point. This leads to $\bar {c}_{\rm{me},PNM} = 0.0025$ and $\bar {c}_{\rm{me},SNM} = 0.02$ for $\varepsilon_\mathrm{fs}$. 
The finite-size effects of different densities are clearly correlated 
while there are insufficient data to learn its correlation structure. 
Here we simply used $0.81~\rm{fm}^{-1}$ ($0.34~\rm{fm}^{-1}$) as 
the correlation length for pure neutron matter (symmetric nuclear matter) 
and assume zero correlation between pure neutron matter and symmetric nuclear matter.

\begin{figure}[p]
\includegraphics[width=0.85\linewidth]{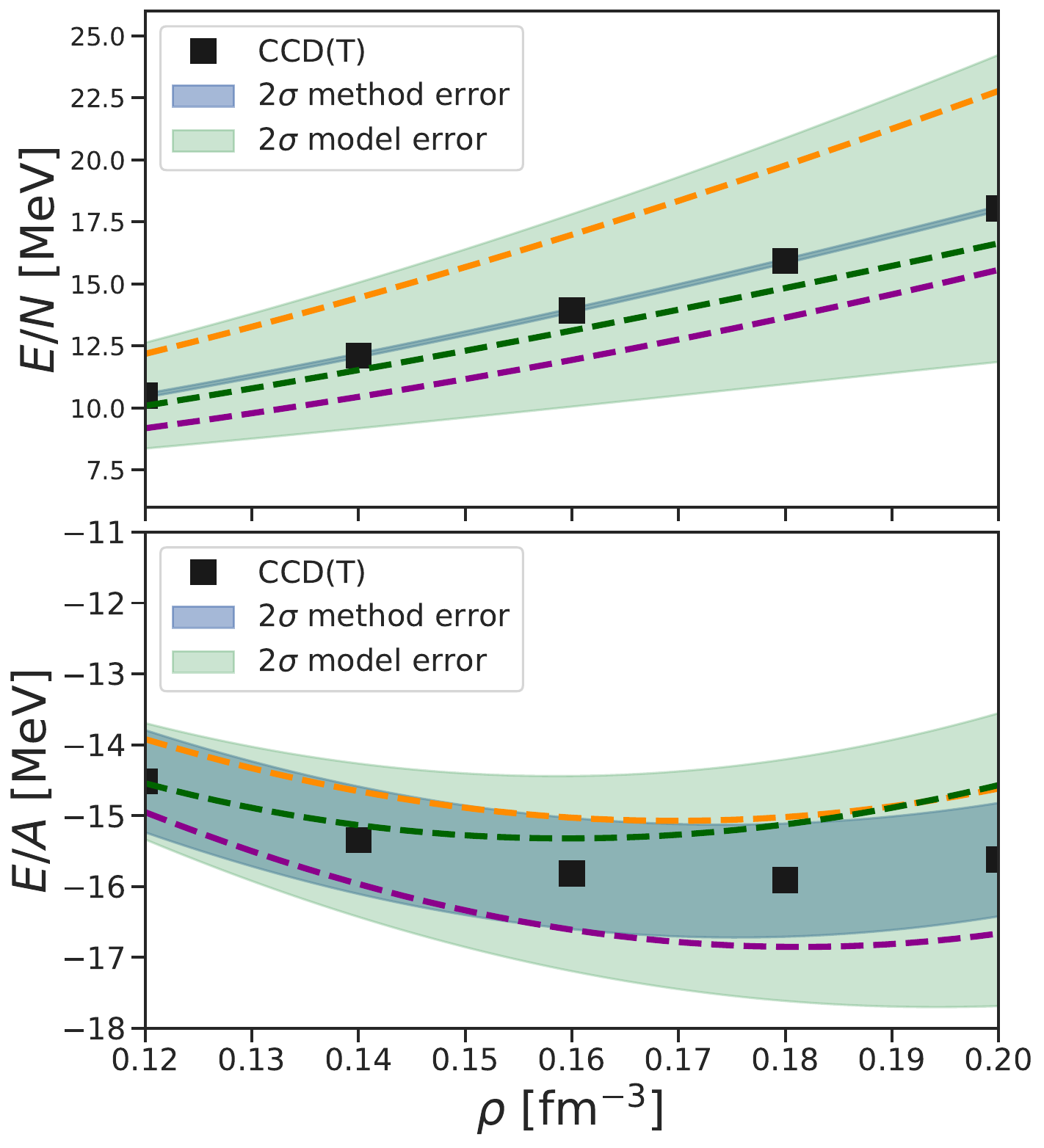}
\caption{ \textbf{Bayesian machine learning error model.}
  The equation of state of pure neutron matter (top) and symmetric nuclear matter 
  (bottom) calculated with
  CCD(T) (black squares) are shown along with the corresponding method
  error (blue shade) and EFT truncation error (green shade) for one
  representative interaction.
  Errors are correlated as a function of density $\rho$ and the dashed orange, green and purple curves illustrate predictions with
  randomly sampled method and model errors drawn from the respective
  multitask Gaussian processes. Correlations extend between pure neutron matter $(E/N)$
  and symmetric nuclear matter $(E/A)$ energies per particle which is represented here by curves in the same colour. 
  Note that the method error is very small in neutron matter due to the small finite-size effect and the small differences between CCD and CCD(T) results (the Pauli principle prevents short-ranged three-neutron correlations).
    \label{fig:EOS_with_error}
}
\end{figure}

Once the model and method errors are
determined, it is straightforward to sample these errors from the
corresponding covariance matrix and produce the equation-of-state predictions using
Eq.~\eqref{eq:predicted_observables} for any given interaction. This
sampling procedure is crucial for generating the posterior predictive
distribution of nuclear matter observables shown in Figure~\ref{fig:ppd_FW}a. CCD(T) calculations
for nuclear-matter equation of state and the corresponding $2\sigma$ credible
interval for method and model errors are illustrated in Extended Data
Figure~\ref{fig:EOS_with_error}. The sampling procedure is made
explicit with three randomly sampled equation-of-state
predictions. Note that even though the sampled errors
for one given density appear to be random, the multitask Gaussian processes will
guarantee that the sampled equation of state of nuclear matter are smooth and properly
correlated with each other.


\bibliographystylemeth{naturemag}
\bibliographymeth{master,master_stats}

\textbf{\textcolor{blue}{Data availability}} 
Source data for Figs.~\ref{fig:abinitio}, \ref{fig:blobbogram} and \ref{fig:ppd_FW}a,b,c are provideed with this paper. Furthermore, the parameters of the 34 non-implausible interactions that is the final result of Extended Data Fig.~\ref{fig:hm_waves} plus the mean vector and covariance matrix of a multivariate normal distribution that approximates the full posterior predictive distribution shown in Fig.~\ref{fig:ppd_FW}a are also provided.
The data that support the other figures of this study are available from the corresponding author upon reasonable request.

\textbf{\textcolor{blue}{Code availability}} 
The code used to perform the IMSRG calculations is available at https://github.com/ragnarstroberg/imsrg.
Inquiries about other codes used in this work should be adressed to the corresponding author.

\textbf{\textcolor{blue}{Acknowledgements}}
This material is based upon work supported by the Swedish Research
Council grant numbers 2017-04234 (C.F.\ and W.J.), 2021-04507 (C.F.), and 2020-05127 (A.E.), the European Research Council (ERC)
under the European Union's Horizon 2020 research and innovation
programme (grant agreement number 758027) (A.E.\ and W.J.), the U.S.\ Department of
Energy, Office of Science, Office of Nuclear Physics under award
numbers DE-AC02-06CH11357 (S.R.S.), DE-FG02-97ER41014 (S.R.S.), DE-FG02-96ER40963 (T.P.) and DE-SC0018223 (NUCLEI SciDAC-4
collaboration)(G.H., T.P., Z.S.), the
Natural Sciences and Engineering Research Council of Canada under
grants SAPIN-2018-00027 and RGPAS-2018-522453 (J.H.\ and B.H.), and the Arthur
B. McDonald Canadian Astroparticle Physics Research Institute (J.H.\ and B.H.), UK
Research and Innovation grant EP/W011956/1 (I.V.) and Wellcome grant 218261/Z/19/Z (I.V.),
the Deutsche Forschungsgemeinschaft (DFG, German Research Foundation) -- Project-ID 279384907 -- SFB 1245 (T.M.).
TRIUMF receives funding via a contribution through the National Research Council of Canada. Computer time was provided by the Innovative and Novel
Computational Impact on Theory and Experiment (INCITE) programme. This
research used resources of the Oak Ridge Leadership Computing Facility
located at Oak Ridge National Laboratory, which is supported by the
Office of Science of the Department of Energy under contract
No. DE-AC05-00OR22725 (G.H, T.P., Z.S.), and resources provided by the Swedish National
Infrastructure for Computing (SNIC) at Chalmers Centre for
Computational Science and Engineering (C3SE), and the National
Supercomputer Centre (NSC) partially funded by the Swedish Research
Council through grant agreement no.\ 2018-05973, as well as Cedar at WestGrid and Compute Canada.

\textbf{\textcolor{blue}{Author contributions}}
C.F. led the project. G.H., T.P., W.J., and Z.S.\ performed coupled-cluster
computations. A.E., C.F., W.J., and I.V.\ designed and conducted the history-matching runs, A.E., C.F., W.J., S.R.S.,\ and I.V.\
carried out the statistical analysis. B.S.H., T.M., J.D.H., and S.R.S.\ performed in-medium
similarity renormalization group calculations. All authors aided in writing the manuscript.

\textbf{\textcolor{blue}{Competing interests}}
The authors declare no competing interests.

 
\textbf{\textcolor{blue}{Correspondence and requests for materials}}
should be addressed to C.F.

\textbf{\textcolor{blue}{Reprints and permissions information}}
is available at www.nature.com/reprints



%
%

%
 %
\begin{table*}[p]
 \caption[Error assignments and PPD model checking]
 {\textbf{Error assignments, PPD model checking and validation.}
   Experimental target values and error assignments for observables used in the
   fourth wave of the iterative history matching (history-matching observables), for the
   Gaussian likelihood calibration of the final non-implausible samples (calibration
   observables), and for model validation with predicted \Pb{}
   observables and electric dipole polarizabilities
   (validation observables). 
   Energies $E$ (in MeV) with experimental targets from
   Refs.~\citemeth{wang2020, multhauf1975}, point-proton radii $R_\mathrm{p}$ (in fm)
   with experimental targets translated from measured charge
   radii~\citemeth{angeli2013} (see Ref.~\citemeth{carlsson2016} for
   more details).
   For the deuteron quadrupole moment $Q$ (in $e^2\mathrm{fm}^2$) we use
   the theoretical result obtained from the high-precision
   meson-exchange nucleon-nucleon model
   CD-Bonn\citemeth{machleidt2001} as a target with a 4\% error bar.
   Electric dipole polarizability $\alpha_D$ (in fm$^3$) with
   experimental targets from Refs.~\citemeth{birkhan2017, tamii2011}.
   Theoretical model (method) errors are estimated from the EFT
   (many-body) convergence pattern as discussed in the text. These
   theory errors have zero mean except for the excitation energy $E_{2^+}(^{48}\mathrm{Ca})$
   with $\mu_\mathrm{method} = -1$~MeV from estimated triples and
   $E/A(^{208}\mathrm{Pb})$ with $\mu_\mathrm{method} = -0.36$~MeV/$A$
   from $e_\mathrm{max}$ model-space extrapolation. Emulator errors are estimated
   from cross validation. All errors are represented by the estimated
   standard deviation of the corresponding random vartiable:
   $\sigma_\mathrm{exp} = \mathrm{Var} \left[ \varepsilon_\mathrm{exp}
   \right]^{1/2}$,
   $\sigma_\mathrm{model} = \mathrm{Var} \left[
     \varepsilon_\mathrm{model} \right]^{1/2}$, etc. Most of the
   experimental errors are negligible compared to the theoretical ones
   and therefore given as $\sigma_\mathrm{exp} = 0$. We assume
   that all theory errors are parametrization independent. 
   The final model predictions from the PPD described in the text (and shown
   in Figure~\ref{fig:blobbogram}) are summarized by the medians and
   the marginal 68\% credibility regions in the last column.  
\label{tab:error_assignments}%
}
 \begin{center}
   \begin{tabular}{ccccccr}
     \hline
    \multicolumn{7}{c}{History-matching observables} \\
    Observable  & $z$ & $\sigma_\mathrm{exp}$ & $\sigma_\mathrm{model}$
    & $\sigma_\mathrm{method}$ & $\sigma_\mathrm{em}$ & \multicolumn{1}{c}{PPD} \\
    \hline 
     $E(^{2}\mathrm{H})$ &  -2.2246   &     0    & 0.05 & 0.0005     & 0.001\% & $-2.22_{-0.07}^{+0.07}$\\ 
    %
    $R_\mathrm{p}(^{2}\mathrm{H})$ & 1.976    &    0     & 0.005     & 0.0002 &0.0005\% & $1.98_{-0.01}^{+0.01}$ \\
     $Q(^{2}\mathrm{H})$ & 0.27      &      0.01   & 0.003 &   0.0005     &0.001\%  & $0.28_{-0.02}^{+0.02}$ \\
     $E(^{3}\mathrm{H})$ &  -8.4821   &     0	  & 0.17	 & 0.0005         &0.01\% & $-8.54_{-0.37}^{+0.34}$ \\ 
     $E(^{4}\mathrm{He})$ & -28.2957   &   0     & 0.55     &   0.0005       &0.01\%  & $-28.86_{-1.01}^{+0.86}$ \\ 
    $R_\mathrm{p}(^{4}\mathrm{He})$ & 1.455  &    0     & 0.016    &  0.0002         &0.003\% & $1.47_{-0.03}^{+0.03}$ \\
     $E(^{16}\mathrm{O})$ & 127.62     &    0     & 1.0     &0.75              &{0.5\%} & $-126.2_{-2.8}^{+3.0}$ \\
    $R_\mathrm{p}(^{16}\mathrm{O})$ & 2.58   &     0    & 0.03     &   0.01            &     0.5\% & $2.57_{-0.06}^{+0.06}$ \\
    \hline
    \multicolumn{7}{c}{Calibration observables} \\
    Observable  & $z$ & $\sigma_\mathrm{exp}$ & $\sigma_\mathrm{model}$
    & $\sigma_\mathrm{method}$ & $\sigma_\mathrm{em}$ & \multicolumn{1}{c}{PPD} \\
 \hline 
   $E/A(^{48}\mathrm{Ca})$ &  -8.667   & 0 & 0.54 & 0.25  & --- &  $-8.58_{-0.72}^{+0.72}$ \\
   $E_{2^+}(^{48}\mathrm{Ca})$ &  3.83   & 0 & 0.5 & 0.5  & --- & $3.79_{-0.96}^{+0.86}$ \\
   $R_\mathrm{p}(^{48}\mathrm{Ca})$ & 3.39   & 0 & 0.11 & 0.03  & --- & $3.36_{-0.13}^{+0.14}$ \\
    \hline
    \multicolumn{7}{c}{Validation observables} \\
    Observable  & $z$ & $\sigma_\mathrm{exp}$ & $\sigma_\mathrm{model}$
    & $\sigma_\mathrm{method}$ & $\sigma_\mathrm{em}$ & \multicolumn{1}{c}{PPD} \\
     \hline 
   $E/A(^{208}\mathrm{Pb})$ &  -7.867   & 0 & 0.54 & 0.5  & --- & $-8.06_{-0.88}^{+0.99}$ \\
   $R_\mathrm{p}(^{208}\mathrm{Pb})$ & 5.45   & 0 & 0.17 & 0.05  & --- & $5.43_{-0.23}^{+0.21}$ \\
   $\alpha_D(^{48}\mathrm{Ca})$ & 2.07   & 0.22 & 0.06 & 0.1  & --- & $2.30_{-0.26}^{+0.31}$ \\
   $\alpha_D(^{208}\mathrm{Pb})$ & 20.1   & 0.6 & 0.59 & 0.8  & --- & $22.6_{-1.8}^{+2.1}$ \\
 \hline 
 \end{tabular}
\end{center}
\end{table*}
 %
\begin{figure*}
    \centering
    \includegraphics[width=0.48\textwidth]{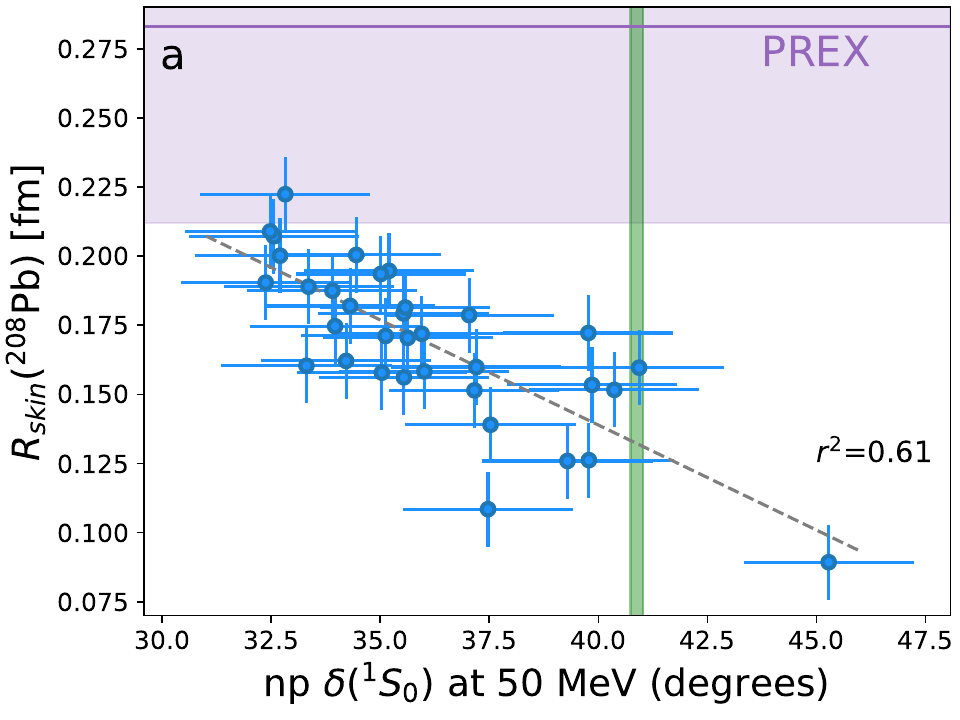}
    \includegraphics[width=0.48\textwidth]{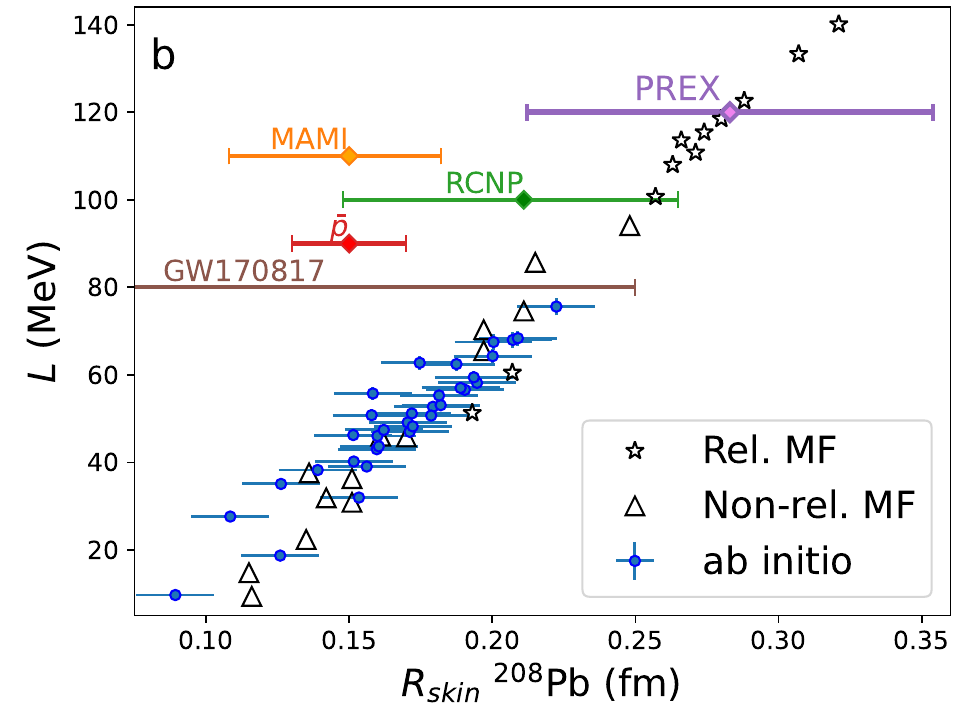}
    \caption[Correlation1]{\textbf{Correlation of \rskinPbbold{} with scattering data and $\boldsymbol{L}$.} \textbf{a} Correlation of computed \rskinPb{} with the proton-neutron $^1S_0$ phase shift $\delta(^1S_0)$ at a laboratory energy of 50~MeV, shown in blue. The error bars represent method and model (EFT) uncertainties. The green band indicates the experimental phase shift\citemeth{navarro2013}, while the purple line (band) indicate the mean result (one-sigma error) of the PREX experiment~\cite{adhikari2021}. The dashed line indicates the linear trend of the {\it ab initio} points with $r^2$ the coefficient of determination. \textbf{b} Correlation of neutron skin \rskinPb{} vs slope of the symmetry energy $L$. Relativistic and non-relativistic mean-field calculations are indicated with open symbols~\citemeth{centelles2010}, while {\it ab initio} results using the 34 non-implausible samples are indicated with filled circles.
      Experimental extractions of \rskinPb{} shown in the figure are from PREX~\cite{adhikari2021}, MAMI~\cite{tarbert2013}, RCNP~\cite{zenihiro2010}, $\bar{p}$~\cite{Trzcinska:2001sy}, and GW170817~\cite{Fattoyev:2017jql}.
      All these results involve modeling input as the neutron skin thickness cannot be measured directly.
      The quoted experimental error bars include statistical and some systematical uncertainties except for Ref.~\cite{Trzcinska:2001sy} that is statistical only and the GW170817 constraint which is a 90 \% upper bound from relativistic mean-field modeling of the tidal polarizability extracted in Ref.~\cite{Fattoyev:2017jql}.
    \label{fig:1S0_and_L_vs_Rskin}}
\end{figure*}
\begin{table*}[p]
 \caption[Predictions for the nuclear matter EOS
     at saturation density and for neutron skins]{\textbf{Predictions for the nuclear equation of state
     at saturation density and for neutron skins.}
     Medians and 68\%, 90\% credible
     regions (CR) for the final PPD including samples from the error models (see
     also Figure~\ref{fig:ppd_FW} and text for details). 
     The saturation density, $\rho_0$, is in ($\mathrm{fm}^{-3}$), the neutron skin
     thickness, \rskinPb{} and \rskinCa{}, in ($\mathrm{fm}$), while the saturation
     energy per particle ($E_0/A$), the symmetry energy ($S$), its
     slope ($L$), and incompressibility ($K$) at saturation density are all in (MeV). 
     Empirical regions shown in Figure~\ref{fig:ppd_FW} are
     $E_0/A = -16.0 \pm 0.5$, $\rho_0 = 0.16 \pm 0.01$, 
     $S=31 \pm 1$, $L=50\pm 10$ 
     and $K=240\pm20$ 
     from Refs.~\cite{lattimer2013}$^,$\citemeth{bender2003, shlomo2006}.
     \label{tab:ppd}
 }
\begin{center}
  \begin{tabular}{crrr}
    \hline
    \multicolumn{4}{c}{Nuclear matter properties} \\
    Observable  & median & 68\% CR & 90\% CR \\
    \hline
    $E_0/A$  & $ -15.2$ &   $[ -16.3, -13.9]$ & $[-17.1,-13.4]$ \\
    $\rho_0$ & $ 0.163$ &   $[ 0.147, 0.176]$ & $ [  0.140,  0.186]$ \\
    $S$  & $ 29.1$ &   $[ 26.8, 31.4]$ & $[ 25.4 33.0]$ \\
    $L$  & $ 50.5$ &   $[ 36.6, 66.3]$ & $[ 23.6, 74.8]$ \\
    $K$  & $ 264$ &   $[ 219, 300]$ & $[202,336]$ \\
    \hline
    \multicolumn{4}{c}{Neutron skins} \\
    Observable  & median & 68\% CR & 90\% CR \\
    \hline
    \rskinCa & $ 0.164$ &   $[ 0.141, 0.187]$ & $ [ 0.123, 0.199]$ \\
    \rskinPb & $ 0.171$ &   $[ 0.139, 0.200]$ & $ [ 0.120, 0.221]$ \\
    \hline\hline
  \end{tabular}
 \end{center}
 \end{table*}

%

\end{document}